\documentclass[aps,prb,floatfix,twocolumn,superscriptaddress,nopacs,longbibliography,epsf]{revtex4-2}

\usepackage[USenglish]{babel}
\usepackage{amsmath,amssymb,amsfonts}
\usepackage{epstopdf}
\usepackage{epsfig} 
\usepackage{graphicx}
\usepackage{subfigure}
\usepackage{import}
\usepackage{xcolor}
\usepackage{url}

\usepackage{tikz}
\usetikzlibrary{tikzmark}

\newlength{\mytextsize}

\newcommand{\JoinUp}[5]{\begin{tikzpicture}[remember picture,overlay,line width=0.05\mytextsize]
    \draw([shift={(#1\mytextsize,#2\mytextsize)}]pic cs:start#5) -- ++(0pt,0.7\mytextsize) -| ([shift={(#3\mytextsize,#4\mytextsize)}]pic cs:end#5);
    \end{tikzpicture}}
\newcommand{\JoinDown}[5]{\begin{tikzpicture}[remember picture,overlay,line width=0.05\mytextsize]
    \draw([shift={(#1\mytextsize,#2\mytextsize)}]pic cs:start#5) -- ++(0pt,-0.7\mytextsize) -| ([shift={(#3\mytextsize,#4\mytextsize)}]pic cs:end#5);
    \end{tikzpicture}}

\usepackage{mathtools}

\allowdisplaybreaks

\usepackage[colorlinks=true,linkcolor=blue,citecolor=red]{hyperref}
\usepackage{changes}

\makeatletter
\renewcommand\tableofcontents{\@starttoc{toc}}
\makeatother
\def\bcen{\begin{center}}
\def\ecen{\end{center}}

\def\a{\alpha}       \def\b{\beta}   \def\g{\gamma}   \def\d{\delta} 
          
\def\k{\kappa}        \def\m{\mu}      
                    \def\s{\sigma}

\def\G{\Gamma}       \def\D{\Delta}

  \def\ie{\mbox{\it i.e.\ }}
\def\=={\equiv}

\def\qed{\raise1pt\hbox{\vrule height5pt width5pt depth0pt}}

\def\cG0{{\cal G}_0} 
\def\cG{{\cal G}}

\def\up{\uparrow} \def\down{\downarrow} 

\def\bk{{\bf k}}

\def\ie{\hbox{\it i.e.\ }} 
\def\ie{\mbox{\it i.e.\ }} \def\=={\equiv}
 
  \def\Tr{{\rm Tr}\,}
 \def\ep0{\epsilon_{p}} \def\ed0{\epsilon_{f}}

\def\be{\begin{equation}}
\def\ee{\end{equation}}

\def\cc{c^{\dagger}}
\def\ca{c^{\phantom{\dagger}}}

\newcommand{\ket}[1]{|{#1}\rangle}

\newcommand{\bra}[1]{\langle{#1}|}
\newcommand{\braket}[3]{\langle{#1}| {#2} |{#3} \rangle}

\newcommand{\quave}[1]{\langle{#1}\rangle}

\newcommand{\tr}[1]{{\rm Tr} \left[ {#1} \right]}

\def\pairc{c^{\dagger}_{i \uparrow} c^{{\dagger}}_{i \downarrow}}
\def\paira{c^{\phantom{\dagger}}_{i \downarrow} c^{\phantom{\dagger}}_{i \uparrow}}

\def\raux{\rho_{\rm aux}}

\usepackage{cancel}

\usepackage{fancyhdr}
\begin{document}

\author{Pasquale Filice}
\affiliation{Dipartimento di Fisica dell'Universit\`a di Pisa, Largo Bruno Pontecorvo 3, I-56127 Pisa,~Italy}
\author{Marco Schir\`o}
\affiliation{JEIP, UAR 3573 CNRS, Coll\`ege de France, PSL Research University, 75321 Paris Cedex 05, France}
\author{Giacomo Mazza}
\affiliation{Dipartimento di Fisica dell'Universit\`a di Pisa, Largo Bruno Pontecorvo 3, I-56127 Pisa,~Italy}

%\title{Non-Hermitian limit of the dissipative dynamics in open BCS superconductors}
\title{Time-dependent Variational Principles for Hybrid Non-Unitary Dynamics: Application to Driven-Dissipative Superconductors}	

%\begin{abstract}
%We study the crossover between the Lindblad and 
%the non-Hermitian (NH) limits of the dissipative dynamics 
%in open supercondutors.
%We show that the NH limit acts as a singular 
%limit of the dissipative dynamics, leading to a sharp 
%modification of the universal approach to the 
%dissipative steady-states. 
%We show that this is due to the fact that, in the NH limit, 
%the dissipative  dynamics preserves the length of the 
%pseudospins. 
%By considering dissipative dynamics with pair losses,
% we show that, which the dynamics in the Lindblad and 
% NH limits converge towards the same steady-state.
% By considering the dynamics with simultaneous pair losses and pumps
% the two dynamics converges to different steady-states.
% Specifically, the steady state reached by the 
% dynamics in the NH limit is different from the infinite 
% temperature state reached in the presence of a finite contribution
% of the jumps.  
%\end{abstract}

\begin{abstract}
We introduce time-dependent variational principles to study the non-unitary 
dynamics of open quantum many-body systems, including dynamics described by 
the full Lindblad master equation, the non-Hermitian dynamics corresponding to the 
no-click limit of the fully post-selected quantum trajectories, and 
the dynamics described by a hybrid Lindbladian with a control parameter 
$\alpha$ which interpolates between the full post-selection and averaging over all 
quantum trajectories.
As an application we study the non-unitary dynamics of a lossy or 
driven-dissipative BCS superconductors, evolving in presence of 
two-body losses and two-body pumps.
We show that the non-Hermitian limit acts as a singular limit of the 
hybrid dissipative dynamics, leading to a sharp modification of 
the universal approach to the driven-dissipative steady-states. 
By considering the dissipative dynamics with pair losses, 
we show that, as the non-Hermitian limit is approached, 
the density dynamics sharply evolves from a universal 
power-law to exponential decay that converges towards a 
quasi-steady plateau characterized by the freezing of the 
particle depletion due to pair losses.
The reached quasi-stationary density increases as a 
function of the dissipation rate highlighting the 
emergence of a non-Hermitian Zeno effect in the lossy dynamics. 
For the driven-dissipative case, we show that, in the non-Hermitian 
 limit, the system gets trapped into an effective negative 
 temperature state, thus skipping the infinite temperature 
 steady-state reached in the presence of finite contribution
 of the quantum jumps.    
 We rationalize these findings in terms of the 
 conservation of the length of the pseudospins which, 
 in the non-Hermitian limit, suppresses the effective 
 single-particle losses and pumps acting on the non-condensed particles.  
\end{abstract}

\maketitle

\section{Introduction}

The dynamics of open quantum systems have attracted the interest of a broad community for several decades now, starting with basic questions on decoherence in quantum mechanics~\cite{weisskopf,wangsness1953thedynamical,breuerPetruccione2007}. In more recent years the experimental developments of controlled quantum simulators based on ultracold atoms in optical lattices and tweezers, superconducting circuits and trapped ions, have focused the interest on dissipation on quantum many-body systems~\cite{fazio2025manybodyopenquantumsystems}. 
Solving the dynamics of a quantum many-body system coupled to an environment represents 
a major computational and conceptual challenge, one for which the development of quantum 
computers could naturally provide an advantage~\cite{trivedi2024quantum,lin2025dissipativepreparationmanybodyquantum}. 

%Solving the dynamics of a quantum many-body system coupled to an environment represents a major computational and conceptual challenge, one for which the development of quantum computers could naturally provide an advantage~\cite{trivedi2024quantum,lin2025dissipativepreparationmanybodyquantum}. 

%Tracing out the environment in the so-called Born-Markov approximation~\cite{lindblad1976onthegenerators,breuerPetruccione2007,quantum_noise_gardiner_zoller}
%leads to a class of open quantum systems whose dynamics is described by the Lindblad master equation.
%Markovian open quantum systems modeled by the Lindblad master equation~\cite{lindblad1976onthegenerators,breuerPetruccione2007,quantum_noise_gardiner}
%represents one of the most important framework.  
%A class of dissipative systems which have attracted large interest are Markovian open quantum systems modelled by a Lindblad master equation~\cite{lindblad1976onthegenerators,breuerPetruccione2007}. This describes the evolution of the reduced density matrix of the system after tracing out the environment.
The dynamics of an open quantum systems is encoded in the non-unitary 
temporal evolution of the reduced density matrix which is obtained 
by tracing the quantum state of the globally isolated system over the environment.
Markovian systems represent an important class of open quantum systems in which 
the non-unitary dynamics of the reduced density matrix is described by the Lindblad 
master equation formalism~\cite{lindblad1976onthegenerators}.
From a microscopic point of view, the Lindblad master equation can be explicitly 
derived upon tracing out a specific environment in the so-called Born-Markov 
approximation~\cite{breuerPetruccione2007,quantum_noise_gardiner_zoller}. 
Alternatively, the Lindblad dynamics naturally emerges in the context of the 
monitored dynamics where it can be obtained as an unraveling over stochastic 
quantum trajectories describing the conditional evolution of the system under 
a series of measurement outcomes (quantum jumps)~\cite{dalibard1992wavefunction,plenio1998quantum,wiseman2009quantummeasurementand}.
%The system evolves with a deterministic non-Hermitian 
%Hamiltonian (NHH) until a stochastic measurement event (quantum jump) projects 
%the wavefunction in a post-measured state.
The full post-selection of quantum trajectories with no jumps, 
also known as the no-click limit~\cite{ashida2020review},
represents a specific limit of the monitored dynamics 
in which the dynamics evolves accordingly to a purely deterministic NHH.
Non-Hermitian (NH) quantum many-body systems have been the focus of several 
theoretical and methodological works~\cite{lee2014heralded,lee2014entanglement,lee2016anomalous,PhysRevLett.121.203001,
biella2021manybodyquantumzeno,gopalakrishnan2021entanglementandpurification,
yamamoto2022universal,hu2023nontrivial,
legal2023volumetoarea,turkeshi2023entanglementandcorrelation,Meden_2023,yu2024nonhermitian,soares2024nonunitary,soares2025symmetriesconservationlawsentanglement}.

A natural interpolation between the full Lindblad evolution and the 
no-click limit is obtained by introducing hybrid Liouvillians characterized by 
a control parameter that weights the quantum jumps contribution on the 
deterministic (unconditional) dynamics.
Hybrid Lindbladians admit natural physical interpretations in term of the 
monitored dynamics such as, for example, a partial post-selection of the 
measurement outcomes~\cite{wiseman2009quantummeasurementand,minganti2020hybrid},
or in terms of the full-counting statistics associated with the number of clicks~\cite{garrahan2010thermodynamics,gupta2024quantum}.
In this context, important questions concern the role of quantum jumps 
for both the conditional and the unconditional dynamics, such 
as, for example, the extent to which the fully post-selected 
deterministic NH evolution can be representative of the full 
Lindblad dynamics or the quantum trajectories~\cite{legal2024entanglement}.

% An intermediate case between full Lindblad evolution and no-click limit is obtained within the so called hybrid Lindbladian, characterized by a control parameter that allows to assess the role of quantum jumps on the unconditional dynamics. 
%In addition to providing a natural interpolation the hybrid Lindbladian formalism admits natural physical interpretations, arising for example when discussing the full counting statistics associated to the number of clicks~\cite{garrahan2010thermodynamics,gupta2024quantum}. Alternatively, one can see the hybrid dynamics as arising from partial post-selection of measurement outcomes or post-selection under imperfect detectors~\cite{wiseman2009quantummeasurementand,minganti2020hybrid}. 
In recent years, several methods have been developed to tackle open quantum many-body 
systems in various setups~\cite{weimer2021simulation}. Examples include diagrammatic 
and renormalization group approaches~\cite{Sieberer_2016}, tensor networks~\cite{cui2015variational,landa2020multistability,kilda2021onthe,mckeever2021stable,shirizly2024dissipative,
Daley2014_quantum_trajec_review,sander2025largescalestochasticsimulationopen}, 
semiclassical methods~\cite{deuar2021fully,naga02021su3}, Faber 
polynomials~\cite{soares2024nonunitary}, as well as neural network states~\cite{Hryniuk2024tensornetworkbased,vicentini2019variational,nagy2019variational,luo2022autoregressive}. 

In the context of the non-equilibrium dynamics of quantum many-body systems, 
time-dependent variational methods are particularly appealing for their physical 
transparency and flexibility. For example, in the unitary case, they have been successfully 
used to describe the dynamics of both weakly and strongly correlated electrons~\cite{schiro_fabrizio_2010,fabrizio2013_timedepedentGA,mazza_electronic_transport_2015,giacomo_temperonics2021}.
So far, variational approaches to dissipative systems focused almost 
exclusively on the steady-state properties~\cite{weimer2015variational},
whereas time-dependent variational Lindbald dynamics have been 
recently discussed within phase space methods~\cite{eeltink2023variationaldynamicsopenquantum} or 
in the context of dissipative superconductors~\cite{mazza_schiro_PRA_2023}
and impurity models~\cite{yifan2025_variational}.

In this work we introduce a general class of time-dependent variational 
principles to describe dynamics in various types of open quantum systems problems, 
including the Lindblad evolution, its fully post-selected version described 
by a NHH or the dynamics under partial post-selection, described by hybrid Liouvillians.  
The variational principles represents the extension to the non-unitary case 
of the action-based Dirac-Frenkel variational principle for unitary dynamics 
and allows to derive effective Liouvillians taking into account the 
many-body nature of the dissipators.
We apply our formalism to driven-dissipative BCS superconductors~\cite{yamamoto2021collective,YamamotoEtAlPRL19,mazza_schiro_PRA_2023,nava2023lindblad,nava2024dissipation}, relevant, for example, for the dynamics of ultracold 
superfluid fermions in presence of particle losses~\cite{PhysRevLett.130.063001}
or for quantum analogue simulators in cavity QED setups~\cite{Young_dynmicalPhasesBCScavity2024}.
We discuss the crossover of the non-unitary dynamics from the Lindbald 
to the fully post-selected NH limit in the case of two-body losses and 
the driven-dissipative case of simultaneous two-body pumps and losses. 
We show that, due to the suppression of effective single 
particle dissipation, the no-click limit represents a singular 
limit of the non-unitary dynamics in which both the universal approach 
to the steady state and the steady state itself are sharply modified.

The manuscript is organized as follows. In Sec.~\ref{sec:open}, we introduce the basic formalism of open quantum systems that will be used throughout this work. 
In Sec.~\ref{sec:non_unitary_TDVP}, we introduce the time-dependent variational principles 
for the hybrid non-unitary dynamics of dissipative fermions. 
In Sec.~\ref{sec:application}, we show the detailed application of the time-dependent variational 
principles to the driven-dissipative BCS superconductors, and in Sec.~\ref{sec:results}, 
we present our results for the hybrid non-unitary dynamics in the presence of both 
two-body losses and pumps. 
Finally, in Sec.~\ref{sec:conclusions} we summarize the results of 
this work and draw our conclusions.

\section{Hybrid Non-Unitary Dynamics of Open Quantum Systems}\label{sec:open}
In this section we recall the basics of the dynamics of open quantum systems, from Lindblad evolution to the non-Hermitian dynamics~\cite{fazio2025manybodyopenquantumsystems}. We focus on the dynamics of open quantum many-body systems described by a Hamiltonian $\mathcal{H}$ 
and a set of independent environments.  In practice, we will always assume a Markovian description 
of the environments. In this limit, the non-unitary dynamics of the system, 
obtained upon tracing out the environments, is naturally modeled by the Lindblad master equation
\begin{equation}\label{eq:lindblad_0.0}
    d\rho(t)=-idt\left[\mathcal{H},\rho\right]+dt\sum_\mu \left( L_\mu\rho(t)L^{\dagger}_\mu-\frac{1}{2} \left\{L^{\dagger}_\mu L_\mu,\rho\right\} \right), 
\end{equation}
where $\rho(t)$ represents the density matrix of the system. 
The first term on the right-hand side of Eq.~\eqref{eq:lindblad_0.0} describes the 
coherent dynamics driven by the Hamiltonian. The second term represents the dissipative (non-unitary)
component of the Lindblad equation defined by a set of jump 
operators $L_{\mu}$ accounting for the effect of the environments on the system. 

The two terms entering the dissipative component of the Lindblad 
equation can be interpreted more clearly by unravelling the dynamics of the reduced density matrix 
into quantum trajectories. 
Quantum trajectories are pure states $\ket{\psi(\xi_t,t )}$ 
that evolve accordingly to the following stochastic Schr\"odinger 
equation (SSE) defined by a set of statistically independent 
Poisson processes 
$\xi_t=\left\{\xi_{\mu,t}\right\}$,
\begin{equation}
\begin{aligned}
d\ket{\psi(\xi_t,t )} 
&= -i dt \left[\mathcal{H}-\frac{i}{2}\sum_\mu \left(L^\dagger_\mu L_\mu-\langle L^\dagger_\mu L_\mu\rangle_t \right)\right]\ket{\psi(\xi_t,t )}  \\
&\quad
+ \sum_\mu\left(\frac{L_\mu}{\sqrt{\langle L^\dagger_\mu L_\mu\rangle}_t}-1\right) d\xi_{\mu,t}\ket{\psi(\xi_t,t )},
\end{aligned}
\label{eq:qjump}
\end{equation}
where $\langle\circ\rangle_t\equiv \langle \psi(\xi_t,t) \vert\circ\vert \psi(\xi_t,t)\rangle$,
$d\xi_{\mu,t} d\xi_{\mu',t} = \d_{\m \m'} d\xi_{\mu,t} $ and ${d\xi_{\mu,t}\in\{0,1\}}$ 
with expectation values
$ E\left[{d \xi_{\mu,t}} \right]  = dt \quave{L^{\dagger}_\mu L_\mu}_t$.
The dynamics of a single quantum trajectory 
breaks down into two steps: a deterministic non-unitary evolution governed by 
the NHH
\begin{align}
    \mathcal{H}_{\rm nH}=\mathcal{H}-\frac{i}{2}\sum_\mu L^\dagger_\mu L_\mu ,
    \label{eq:nhH}
\end{align}
corresponding to the so-called recycling term in Eq.~\ref{eq:lindblad_0.0} written in terms 
of anticommutator, and a series of stochastic quantum jumps at random times (second line of Eq.~\eqref{eq:qjump}), at which the state $\ket{\psi(\xi_t,t)}$ changes discontinuously. 
We note that the stochastic evolution is normalised and state dependent,
as is evident from the expectation values $\quave{L^\dagger_\mu L_\mu}_t$ 
appearing in both the deterministic and the stochastic parts of Eq.~\eqref{eq:qjump},
and in the rates of the Poisson processes.
Given a quantum trajectory, one can define the conditional 
density matrix $\rho_c(\xi_t,t)=\vert \psi(\xi_t,t)\rangle\langle  \psi(\xi_t,t)\vert$.
Considering all possible quantum trajectories, the density matrix averaged 
over all measurement outcomes 
\begin{equation}
    E \left[ \rho(t) \right] \equiv \lim_{N_\xi \to \infty} \frac{1}{N_\xi} \sum_{\xi_t} {\rho_c(\xi_t,t)},
\end{equation}
evolves according to the Lindblad master equation in Eq.~(\ref{eq:lindblad_0.0}). 
On the contrary, if one post-selects the quantum trajectories over the records of 
no clicks, the dynamics is deterministic and 
completely governed by $\mathcal{H}_{\rm nH}$. 

An intermediate case between the Lindblad evolution and no-click limit is obtained 
by defining the temporal evolution  of the density matrix accordingly to a modified
Lindbladian parametrized by a real parameter $\a \in [0,1]$
\begin{equation}\label{eqn:hybridLindblad}
    d\rho_{\alpha}(t)= dt {\cal L}_\a[\rho_{\a}(t)]
\end{equation}    
with 
\begin{equation}
{\cal L}_\a 
= 
-i \left[\mathcal{H},\rho_{\alpha}(t)\right] + \d {\cal L}_{\a}[\rho_\a(t)]
\label{eq:liouvillian_alpha}
\end{equation}
and dissipator
\begin{equation}
\d {\cal L}_\a 
= \sum_\mu \left(\alpha L_\mu\rho_{\alpha}(t)L^{\dagger}_\mu-\frac{1}{2} \left\{L^{\dagger}_\mu L_\mu,\rho_{\alpha}\right\} \right).
\label{eq:dissipator_alpha}
\end{equation}
The hybrid dissipator in Eq.~\eqref{eq:dissipator_alpha} interpolates 
between the full Lindblad dynamics for $\alpha=1$ and the post
-selected no-clicks non-Hermitian evolution for $\alpha=0$. 
For later convenience, we notice that 
the hybrid dissipator can be equivalently 
written in the form of a full 
Lindblad dissipator with renormalized coupling 
supplemented by a purely non-Hermitian contribution
\begin{equation}
\d {\cal L}_{\a} = \a \d {\cal L}_{1} +\frac{1}{2} (\a-1) 
\sum_{\mu} \left\lbrace L^{\dagger}_{\mu} L^{\phantom{\dagger}}_{\mu}, \rho_\a \right\rbrace,
\end{equation} 
where $\d {\cal L}_1 \equiv \d {\cal L}_{\a=1}$. 

Equations~\eqref{eqn:hybridLindblad}-\eqref{eq:dissipator_alpha} define the 
general form of hybrid Liouvillians considered in this work.
We notice that, for any value $  0 \leq \a  < 1$, namely for all cases except for the Lindblad 
one $\a=1$, the dynamics governed by the hybrid Liouvillian does not conserve the norm. 
We therefore include an explicit normalization in the definition of the 
expectation values, \ie
\begin{equation}
\quave{\hat{O}}_{\a}(t) \equiv 
\frac{\Tr \left( \rho_{\a}(t) \hat{O} \right)}{\Tr \left( \rho_{\a} (t) \right)}
\label{eq:Oalpha_def}
\end{equation}
%\ms{This in fact comes naturally from the interpretation as measurment problem (perhaps to elaborate)?.}\gm{fix this by MS}
This in fact comes naturally from the interpretation of quantum trajectories  as measurement problem:
after any non-unitary step, either a click or a no-click, the state is properly normalized as enforced both for the quantum jump term 
as well as for the non-Hermitian evolution in Eq.~\eqref{eq:qjump}. Assuming the hybrid Liouvillian evolution Eq.~\eqref{eqn:hybridLindblad},  
 the equations of motion for the expectation values $\quave{\hat{O}}_\a$ read
\begin{equation}
\begin{split}
 \frac{d \quave{\hat{O}}_\a}{dt}
&= -i \quave{\left[ \hat{O}, {\cal H}  \right]}_\a \\
& 
+ \frac{\a}{2} \sum_{\mu} \left( 
\quave{L^{\dagger}_{\mu} \left[ \hat{O},L_{\mu}^{\phantom{\dagger}} \right] }_\a 
-
\quave{ \left[ \hat{O},L_{\mu}^{{\dagger}} \right] L^{\phantom{\dagger}}_{\mu} }_\a 
\right)   \\
& + 
\frac{\a-1}{2} 
\sum_\mu
\left( 
\quave{\left\lbrace L_{\mu}^{\dagger} L_{\mu}^{\phantom{\dagger}} \hat{O} \right\rbrace}_\a
- 
2
\quave{L_{\mu}^{\dagger} L_{\mu}^{\phantom{\dagger}}}_\a \quave{ \hat{O} }_\a
\right),
\end{split}
\label{eq:EOM_Oalpha}
\end{equation}
where the last term comes from the normalization introduced 
in Eq.~\eqref{eq:Oalpha_def}.
We observe that the same equations of motions can 
be equivalently obtained by supplementing the hybrid 
Liouvillian with non-linear term 
\begin{equation}
\bar{\cal L}_{\a} = {\cal L}_{\alpha} - \rho_\a (\a-1) \sum_{\mu}
\quave{L_{\mu}^{\dagger} L_{\mu}^{\phantom{\dagger}}}_\a,
\label{eq:norm_conserving_hybrid_L}
\end{equation}
such that $\tr{\bar{\cal L}_\a} = 0$
the resulting dynamics is norm conserving.
By taking the $\a=0$ limit, it is immediate to check that the 
non-linear term in Eq.~\eqref{eq:norm_conserving_hybrid_L} is 
equivalent to the normalization introduced for the 
stochastic Schr\"odinger equation in the no-click limit.

\section{Time-Dependent Variational Principles}
\label{sec:non_unitary_TDVP}
In this section, we introduce a class of 
time-dependent variational principles for the hybrid non-unitary 
dynamics. 
We consider variational principles defined by the application of 
stationary conditions on suitably defined actions. 
In the case of unitary dynamics, the variational dynamics is expressed 
by the Dirac-Frenkel variational principle (DFVP), defined by the action 
\begin{equation}
{\cal S}_{\rm DF} = \int d t \braket{\Psi(t)}{i \partial_t - H}{\Psi(t)} 
\label{eq:DFVP}
\end{equation}
and the stationary condition $\d {\cal S}_{\rm DF} = 0$.
Eq.~\eqref{eq:DFVP} can be viewed as the least action definition 
of the time-dependent Schr\"odinger equation.
Here, we extend this concept to the non-unitary case. 
We start from the full Lindbald dynamics $\a=1$
and introduce an action ${\cal S}[\rho_{v},\rho_{\rm aux}]$ 
which is a functional of a variational state $\rho_v(t)$ and 
of an auxiliary state $\rho_{\rm aux}(t)$
\begin{equation}
{\cal S}[\rho_{v},\rho_{\rm aux}] = 
\int dt \tr{\rho_{\rm aux} \left( \partial_t \rho_{v}(t) - {\cal L}_1[\rho_v(t)]  \right)}.
\label{eq:action_DMVP}
\end{equation}
Starting from the action Eq.~\eqref{eq:action_DMVP}, 
we define the density matrix variational principle (DMVP)
by requiring the stationarity of the action with respect  
to the auxiliary state, \ie 
\begin{equation}
\frac{\d {\cal S}[\rho_{v},\rho_{\rm aux}]}{\d \rho_{\rm aux}} = 0.
\label{eq:DMVP_original}
\end{equation}
Analogously to the DFVP, the DMVP in Eq.~\eqref{eq:DMVP_original} 
can be viewed as a least action definition of the Lindblad master equation.
The choice of the auxiliary state is completely 
arbitrary and it is formally introduced to take the variation of 
the action.
In particular, the choice of $\raux$ does not affect the variational 
freedom of the problem. On the contrary, the variational freedom is 
constrained by the the choice of the specific variational density 
matrix $\rho_v(t)$ that, for an arbitrary $\raux$, defines the 
action associated with the variational state.
We notice that, upon introducing the vectorization (or superfermion) formalism, where the Lindblad superoperator becomes a non-Hermitian matrix 
acting on vectorized density matrices~\cite{PhysRevB.89.165105,PhysRevB.109.125125},
 the DMVP can be recast in a similar form to the DFVP
\begin{equation}
{\cal S} = \int dt \left \langle \left \langle {\rho}_{aux} \right. \right| \partial_t - \tilde{\cal L} \left.\left| \rho_v \right \rangle \right \rangle 
\label{eq:DMVP_superoperator}
\end{equation}
where the symbol $| \rangle \rangle$ indicates vectors in the product Hilbert space
and  $\tilde{\cal L}$  is the superoperator definition of the Liouvillian.
The DMVP is therefore obtained by taking the variation with 
respect to the bra-vector 
${\d {\cal S}}/{\d \left \langle \left \langle {\rho}_{aux} \right. \right| } = 0$.

The DMVP can be readily extended to the hybrid case $\a \neq 1$.
In this case, the variational principle must include also the state normalization.
Following the discussion of the previous section, the 
are two equivalent ways of extending the DMVP to the case of 
the hybrid Liouvillian by keeping into account the state normalization.
In the first case, one uses the original, \ie non norm-conserving, 
hybrid Liouvillian Eq.~\eqref{eq:liouvillian_alpha} and 
explicitly normalizes all the traces with respect to 
the trace of the variational density matrix.
Specifically, the DMVP becomes
\begin{equation}
\frac{\d}{\d \rho_{\rm aux}} \int 
\frac{\Tr \left[ \rho_{\rm aux} ( \partial_{t} \rho_v(t) -  {\cal L}_{\a}[\rho_v(t)] ) \right]}
{\tr{\rho_v(t)}} = 0.
\label{eq:DMVP_hybrid_normalized}
\end{equation} 
Equivalently, one may consider the non-linear norm conserving hybrid 
Liouvillian, Eq.~\eqref{eq:norm_conserving_hybrid_L}, and write
\begin{equation}
\frac{\d}{\d \rho_{\rm aux}} 
\int d t \tr{\rho_{\rm aux} \left(  \partial_t \rho_{v}(t) - \overline{\cal L}_{\a}[\rho_v(t)]  \right)} 
= 0.
\label{eq:DMVP_hybrid_nonlinear}
\end{equation}
Obviously, Eqs.~\eqref{eq:DMVP_hybrid_normalized}-\eqref{eq:DMVP_hybrid_nonlinear}
reduce to Eq.~\eqref{eq:DMVP_original} for $\a=1$.

\subsection{Non-Hermitian Dirac-Frenkel Variational Principle}
We now consider the no-click limit, $\a=0$, of the variational 
non-unitary dynamics. 
In this limit, the variational dynamics can be equivalently expressed 
in terms of a non-Hermitian extension 
of the DFVP (NH-DFVP). The NH-DFVP reads
\begin{equation}
\d \int dt \braket{\Psi(t)}{i \partial_t - {\cal H}_{\rm nH}}{\Psi(t)} 
+ \mu(t) \left(\braket{\Psi(t)}{}{\Psi(t)} -1\right) = 0,
\label{eq:nH-WFVP}
\end{equation}
where $\mu(t)$ is a time-dependent Lagrange 
multiplier ensuring the normalization of the state.
Taking the variation with respect to both the state 
and the Lagrange parameter one obtains 
$\mu(t) = \frac{i}{2} \sum_{\mu} \braket{\Psi(t)}{L^{\dagger}_{\mu} L_{\mu}}{\Psi(t)}$
matching the definition of the norm-conserving 
NHH evolution in the no-click limit of the SSE, see Eqs.~\eqref{eq:qjump}. 
It is immediate to check that, by constructing a time-dependent mixed state 
\begin{equation}
\rho(t) = \sum_{n} p_n \ket{\Psi_n(t)} \bra{\Psi_n(t)}
\end{equation}
with $p_n>0$ and $\sum_n p_n=1$,
and upon extending the definition of the action for 
the NH-DFVP as 
\begin{equation}
\begin{split}
{\cal S}_{NH} = \sum_n p_n\int dt &
\braket{\Psi_n(t)}{i \partial_t - {\cal H}_{\rm nH}}{\Psi_n(t)} \\
&+ \mu(t) \left(\braket{\Psi_n(t)}{}{\Psi_n(t)} -1\right),
\end{split}
\end{equation}
the stationary condition $\d {\cal S}_{NH} =0$ leads to the same 
variational dynamics determined by the DMVP for $\a=0$.
We notice that the NH extension of the DFVP can applied in 
combination with the method of quantum-trajectories to obtain 
a variational description of the non-unitary dynamics governed by 
the Stochastic Scrh\"odinger Equation.

%We end this section by noticing that the NH-DFVP for pure states 
%can be straightforward applied to the variational description of the 
%stochastic dynamics governed by the SSE.

\subsection{Example: Variational Gaussian States}
\label{sec:example_gaussian}
We now show a practical implementation of the above variational principles
in the case of Gaussian variational density matrices
$\rho_v(t) = \rho_0(t).$ 
In this case, it is possible to formally calculate the action 
without an explicit definition of the auxiliary state and 
by retaining the full functional dependence of the action on $\raux$.
We consider a Liouvillian built from a generic many-body 
Hamiltonian of the form
\begin{equation}
{\cal H} = \sum_{a b} t_{a b} \cc_a \ca_{b}
+ \frac{1}{2} \sum_{a b c d} U_{a b c d} 
\cc_a \cc_{b} \ca_{c} \ca_{d}
\end{equation}
and a many-body jump operator describing two-particle losses
\begin{equation}
L = \sum_{a b} \sqrt{\k_{a b}} \ca_{a} \ca_{b}.
\label{eq:2p_loss_jump}
\end{equation}
Here, $\cc_{a}$ and $\ca_{a}$ are  
fermionic creation and annihilation operators, 
and we used latin indexes to label generic quantum numbers 
associated  with the latter.
$t_{ab}$ and $U_{abcd}$ are, respectively, the hopping and 
the interaction matrix elements, and $\sqrt{\k_{ab}}$ are 
the dissipation rates in the $ab$ channel.

We start from the limit $\a=1$ and compute the trace 
$\Tr  \left( \rho_{\rm aux} {\cal L}_1[\rho_0(t)] \right) .$
As shown in the appendix~\eqref{app:DMVP_Guassian}, the gaussianity of the state 
allows to express this 
trace in terms of an effective Hartree-Fock single-particle Liouvillian 
\begin{equation}
\Tr  \left( \rho_{\rm aux} {\cal L}_1[\rho_0(t)] \right)
= 
\Tr  \left( \rho_{\rm aux} {\cal L}_{HF}[\rho_0(t)] \right)
\label{eq:trraux_HF}
\end{equation}
with 
\begin{equation}
{\cal L}_{HF}[\rho_0] = -i \left[ {\cal H}_{HF}[\rho_0],\rho_0 \right] 
+ \d {\cal L}_{HF}[\rho_0].
\end{equation}
Here, ${\cal H}_{HF}[\rho_0(t)]$ is the usual time-dependent 
Hartree-Fock single-particle Hamiltonian
\begin{equation}
{\cal H}_{HF} =
\sum_{ab} t_{ab} \cc_{a} \ca_{b} 
+ \sum_{abcd} V_{abcd}
\Tr \left( \rho_0(t) \cc_a \ca_d \right)
\cc_{b} \ca_{c}
\label{eq:Hhf_generic}
\end{equation}
with $V_{abcd} \equiv U_{abcd}-U_{abdc}$, and 
$\d {\cal L}_{HF}[\rho_0]$ is an effective Hartree-Fock dissipator
which reads
\begin{equation}
\begin{split}
\d {\cal L}_{HF}
=& 2\sum_{abcd} {\G}_{abcd}
\Tr \left(\rho_0 (t) \cc_{a} \ca_{d} \right)
\d {\cal L}^{cb}[\rho_0] 
\end{split},
\end{equation}
where $\d {\cal L}^{cb}[\rho_0]$ is a single-particle loss 
dissipator
\begin{equation}
\d {\cal L}^{cb}[\rho_0]
=
\ca_{c} \rho_0 \cc_{b} - \frac{1}{2} \left\lbrace \cc_{b} \ca_{c},\rho_0 \right\rbrace,
\end{equation}
and
${\G}_{abcd} \equiv \sqrt{\k_{ba} \k_{cd}} - 
\sqrt{\k_{ba} \k_{dc}} $.
We notice that, due to the many-body nature of the original jump 
operators, the effective Hartree-Fock single-particle dissipator 
is characterized by effective couplings which depend 
on the variational state itself.

By taking the variation of the action defined by Eq.~\eqref{eq:trraux_HF} 
with respect to the auxiliary density matrix $\rho_{\rm aux}$, 
we obtain the effective Liouvillian for the variational density 
matrix
\begin{equation}
\partial_t \rho_0(t)
= - i \left[{\cal H}_{HF}[\rho_0],\rho_0 \right]
+ \d {\cal L}_{HF}[\rho_0].
\end{equation}
Let us now consider the $\a<1$ case.
Using the normalized variational principle in Eq.~\eqref{eq:DMVP_hybrid_normalized}, 
the calculation of the traces on the normalized state is carried out exactly 
as before, and the effective hybrid Liouvillian for the Gaussian state reads
\begin{equation}
\begin{split}
 \partial_t &\rho_{0,\a}(t)
= - i 
\left[ {\cal H}_{HF}[\rho_{0,\a}],\rho_{0,\a}\right]
+ \a \d {\cal L}_{HF}[\rho_{0,\a}] \\
&+ (\a-1) \sum_{a b c d} {\G}_{abcd}
\Tr \left(\rho_{0,\a} (t) \cc_{a} \ca_{d} \right)
\left\lbrace \cc_{b} \ca_{c},\rho_{0,\a} \right\rbrace.
\end{split}
\label{eq:effective_liouvillian_DMVP_normalized}
\end{equation}
On the other hand, by using the norm-conserving variational principle
Eq.~\eqref{eq:DMVP_hybrid_nonlinear}, the effective hybrid Liouvillian 
becomes 
\begin{equation}
\begin{split}
 \partial_t &\overline{\rho}_{0,\a}(t)
= - i 
\left[ {\cal H}_{HF}[\overline{\rho}_{0,\a}],\overline{\rho}_{0,\a}\right]
+ \a \d {\cal L}_{HF}[\overline{\rho}_{0,\a}] \\
&+ (\a-1) \sum_{a b c d} {\G}_{abcd}
\left[ \Tr \left(\overline{\rho}_{0,\a} (t) \cc_{a} \ca_{d} \right)
\left\lbrace \cc_{b} \ca_{c},\overline{\rho}_{0,\a} \right\rbrace \right. \\ 
&  \left. - 2(\a-1)\overline{\rho}_{0,\a}(t)
\Tr \left(\overline{\rho}_{0,\a} (t) \cc_{a} \ca_{d} \right)
\Tr \left(\overline{\rho}_{0,\a} (t) \cc_{b} \ca_{c} \right) \right]. 
\end{split}
\label{eq:effective_liouvillian_DMVP_normconserving}
\end{equation}
On can readily check the equivalence between Eq.~\eqref{eq:effective_liouvillian_DMVP_normalized} 
and Eq.~\eqref{eq:effective_liouvillian_DMVP_normconserving}
by computing the equations of motions 
for the normalized observables 
\begin{equation}
\quave{\hat{O}}_\a = \frac{\Tr \left( \rho_{0,\a} \hat{O} \right) }{\Tr\left( \rho_{0,\a} \right)} =
\Tr \left( \overline{\rho}_{0,\a} \hat{O} \right).
\end{equation}
%We end this section by stressing the fact that, despite the 
%specific example on Gaussian states, the above variational 
%procedure is completely general and can be applied to other 
%variational states upon suitable choices of the auxiliary 
%density matrix used to impose the stationary condition.

\section{Driven-Dissipative dynamics in BCS Superconductors}\label{sec:application}
We now discuss the application of the time-dependent variational 
principle to the driven-dissipative dynamics in BCS superconductors.
We consider a system of spinful fermions hopping on a lattice with local pairing interaction as described by the attractive Hubbard model 
\begin{align}
{\cal H}=\sum_{ij \s} t_{ij} \cc_{i \s} \ca_{j\s} - |U| \sum_i n_{i \up} n_{i \downarrow}
\label{eq:Hhubbard_attractive}
\end{align}
where $-|U|$ is the attraction and the $t_{ij}$ 
the hopping amplitude between sites $i$ and $j$.
We consider a periodic lattice with $t_{ij} = t_{(i-j)}$, and 
assume that the hopping matrix elements give rise 
to a band of width $W$ characterized by flat-density density of states.
 
At equilibrium, the competition between electron hopping and local 
pairing represents the simplest description of the crossover between 
weak and strong coupling superconductivity which is 
relevant in the contexts of high-temperature superconductors 
and BCS to BEC superfluidity
~\cite{bourdel2004experimental,CHEN20051,biss2022excitation,
toschi2005BCSBEC,mazza_interface_BCSBEC_2021}. 
In out of equilibrium conditions, the unitary dynamics is characterized by 
a series of dynamical phase transitions arising in various quench  protocols~\cite{BarankovLevitovSpivakPRL04,BarankovLevitovPRL06,YuzbashyanEtAlPRB05,YuzbashyanEtAlPRL06,mazza2017fromsudden,seibold2020nonequilibrium,
ojeda2019fate,mazza2012dynamical,seibold2022adiabatic,collado2205}.  
The effect of two-particle losses has been studied in 
Ref.~\onlinecite{YamamotoEtAlPRL19} by approximation of the full 
Lindbald by its NH limit, and by the determination of the spectrum of 
the resulting NHH.
Ref.~\onlinecite{yamamoto2021collective} focused on the full 
Lindblad dissipative dynamics by means of a specific 
decoupling of the interaction which leads to an effective unitary 
dynamics with complex pairing interaction.
Later on, it was showed that the application of the DMVP to the 
full Linbladian, see Eq.~\eqref{eq:DMVP_original}, leads to a much 
richer dynamics that clearly shows how the non-unitary effects 
discarded in previous works dominates the dissipative 
dynamics~\cite{mazza_schiro_PRA_2023}. 

Here, we consider the hybrid driven-dissipative dynamics 
by considering both two-particle losses and pumps
\begin{equation}
\d {\cal L}_{\a,\a'} = \d {\cal L}_{\a}^{\rm loss} + \d {\cal L}_{\a'}^{\rm pump}
\label{eq:dissipator_loss_pumps}
\end{equation}
with 
\begin{equation}
\d {\cal L}_{\a}^{X} = 
 \sum_i \left(\alpha L_i^{X} \rho_{\alpha}(t){L_i^X}^{\dagger}-\frac{1}{2} 
\left\{ {L_i^X}^{\dagger} L_i^{X},\rho_{\alpha}\right\} \right)
\end{equation}
for $X \in ({\rm loss,pump})$, and 
\begin{equation}
L_i^{\rm loss} = \sqrt{2 \G} \ca_{i \down} \ca_{i \up}
\end{equation}
and 
\begin{equation}
L_i^{\rm pump} = \sqrt{2 P}  \cc_{i \up} \cc_{i \down}.
\end{equation}
In the definition of the dissipator, Eq.~\eqref{eq:dissipator_loss_pumps}, 
we explicitly highlight that, in principle, one may 
consider different parameters $\a$ and $\a'$ for the hybrid 
non-unitary dynamics of losses and pumps.
In the following we consider only the case $\a=\a'$.

%The resulting dissipative dynamics does not conserve the total number of particles. In absence of any driving term to counterbalance the loss of particles into the environment the system evolves at long times towards the zero density limit. We note that two-body losses conserve instead the total spin which would prevent from reaching complete depletion~\cite{rosso2021onedimensional}, unless the system is initially prepared in a total singlet state as it is our case here. While the stationary state properties of the model are therefore trivial its depletion dynamics can still reveal intriguing features and give rise to different dynamical regimes, as we are going to discuss.

\subsection{BCS hybrid Liouvillians}
\label{sec:bcs_hybrid_liouvillian}
We derive the equation of motions for the 
hybrid dissipative dynamics by using a BCS 
ansatz for the variational density matrix.
The BCS state is a Gaussian state allowing 
for non zero anomalous contractions, \ie $\Tr \left(\rho_0 \cc_{\up} \cc_{\down} \right) \neq 0$.
We use the normalized version of the variational 
principle, Eq.~\eqref{eq:DMVP_hybrid_normalized}, and 
we constraint the variational state to preserve 
translation invariance and spin unpolarized density.
By including anomalous contractions in the 
calculation of the action, we obtain the 
effective Liouvillian for the BCS state, see appendix~\ref{sec:hybrid_DMVP_pumps}
for details. 
The BCS state evolves with 
an hybrid effective BCS-Liouvillian 
\begin{equation}
\partial_t \rho_{0,\a} = {\cal L}_{BCS,\a}[\rho_{0,a}]
\label{eq:rhodot_BCS_liouvillian_general}
\end{equation}
with 
\begin{equation}
{\cal L}_{BCS,\a} = -i\left[ H_{BCS}(\G,P), \rho_{0,\a}\right] 
+ \d {\cal L}_{BCS,\a}^{\rm loss}+
\d {\cal L}_{BCS,\a}^{\rm pump}.
\label{eq:BCS_liouvillian_general}
\end{equation} 
The BCS Liouvillian is defined by a unitary BCS Hamiltonian, $H_{BCS}(\G,P)$, 
which depends on the pump and loss rates, 
and two effective dissipators which take into account losses and pumps. 
The unitary Hamiltonian $H_{BCS}(\G,P)$ is obtained by 
adding to the standard BCS Hamiltonian two complex valued 
pairing potentials as
\begin{equation}
H_{BCS}(\G,P) = H_{BCS,0}(t) +i (\G-P) \D(t) \sum_i \ca_{i \down} \ca_{i \up}
+{\rm h.c.}~.
\label{eq:HBCS_complex_pairing}
\end{equation}
Here, 
\begin{equation}
\D(t) \equiv \frac{\Tr\left( \rho_{0,\a}(t) \cc_{i \up} \cc_{i \down} \right)}{\Tr \left( \rho_{0,\a} \right)}
\end{equation}
%$\D(t) \equiv \frac{\Tr\left( \rho_{0,\a}(t) \cc_{i \up} \cc_{i \down} \right)}{\Tr \left( \rho_{0,\a} \right)}$
is the superconducting order parameter, and $H_{BCS,0}$ is the 
usual time-dependent BCS Hamiltonian
\begin{equation}
H_{BCS,0} = \sum_{i j \s} t_{ij} \cc_{i \s} \ca_{j \s}
- |U| \D(t) \sum_i \ca_{i \down} \ca_{i \up} + {\rm h.c.}.
\end{equation}
The effective dissipators in loss and pump channels read
\begin{equation}
\begin{split}
\d & {\cal L}_{BCS,\a}^{\rm loss}
= 2 \G \frac{n(t)}{2}
\d {\cal} L_{\a}^{\rm 1-l}[\rho_{0,\a}] + \\
&+2\G(\a-1) \sum_i  
\D(t) \ca_{i \down} \ca_{i \up} \rho_{0,a}
+ \D(t)^* \rho_{0,\a} \cc_{i\up} \cc_{i \down}
\end{split}
\label{eq:alpha_diss_loss_BCS}
\end{equation}
and
\begin{equation}
\begin{split}
\d & {\cal L}_{BCS,\a}^{\rm pump}
= 2 P \left( 1 -\frac{n(t)}{2} \right) 
\d {\cal} L_{\a}^{\rm 1-p}[\rho_{0,\a}] + \\
&+2P(\a-1) \sum_i  
\D(t) \rho_0 \ca_{i\down} \ca_{i \up}
+ \D(t)^* \cc_{i \up} \cc_{i \down} \rho_0
\\ \\ \\
\end{split}
\label{eq:alpha_diss_pump_BCS}
\end{equation}
In Eqs.~\eqref{eq:alpha_diss_loss_BCS}-\eqref{eq:alpha_diss_pump_BCS}
$\d {\cal L}_{\a}^{\rm 1-l}$ and $\d {\cal L}_{\a}^{\rm 1-p}$ 
represent, respectively, hybrid single-particle pump dissipators
\begin{equation}
\d{\cal L}_{\a}^{\rm 1-l} = 
\sum_{i\s} \left( \a \ca_{i\s} \rho_{0,\a} \cc_{i \s} -\frac{1}{2} 
\left\lbrace \cc_{i \s} \ca_{i \s},\rho_{0,\a} \right\rbrace \right)
\end{equation}
and
\begin{equation}
\d{\cal L}_{\a}^{\rm 1-p} = 
\sum_{i\s} \left( \a \cc_{i\s} \rho_{0,\a} \ca_{i \s} -\frac{1}{2} 
\left\lbrace \ca_{i \s} \cc_{i \s},\rho_{0,\a} \right\rbrace \right),
\end{equation}
and $n(t) = \sum_\s 
\frac{\Tr \left(\rho_{0,\a} \cc_{i \s} \ca_{i \s} \right)}{\Tr \left( \rho_{0,\a} \right)}$
is the total particle density. 

The loss and pump contributions to the effective Liouvillian are 
related by the particle-hole transformation $\ca_{i \s} \to \cc_{i \s}$.
In particular, under particle-hole, 
the order parameter and the particle density transform, respectively, 
as $\D \to -\D^*$ and $n \to 2-n$.  
It is important to observe that, in Eqs.~\eqref{eq:alpha_diss_loss_BCS}-\eqref{eq:alpha_diss_pump_BCS}, the single-particle dissipators 
are characterized by effective dissipation rates which are 
proportional to the number of particles (losses) 
and on the number of holes (pumps).
This fact is due to the many-body nature of the jump 
operators and indicates that the two-body losses (or
the two-body pumps) get more and more disfavored as 
the system get completely depleted (or completely filled).

In the limit $\a=0$, the effective Liouvillian reduces to 
a non-hermitian BCS (NH-BCS) Hamiltonian
\begin{equation}
\left.{\cal L}_{BCS,\a}\right|_{\a=0} = -i H_{{\rm nH}-BCS} \rho_0 + i \rho_{0} H_{{\rm nH}-BCS}^{\dagger}
\end{equation}
with 
\begin{equation}
\begin{split}
H_{{\rm nH}-BCS}  &= H_{BCS}(\G,P)  - i \frac{n}{2} \G \sum_{i \s} \cc_{i \s} \ca_{i \s} \\
& 
- i \left( 1-\frac{n}{2} \right) P \sum_{i \s} \ca_{i\s} \cc_{i \s}.
\end{split}
\label{eq:BCS_nhh}
\end{equation}
It is immediate to check that the NH-BCS Hamiltonian, Eq.~\eqref{eq:BCS_nhh},
can be also directly obtained from the NH-DFVP Eq.~\eqref{eq:nH-WFVP}
without the use of the auxiliary state $\raux$.

\subsection{Equation of motions for normalized observables}
\label{sec:eom_obs_normalization}
We derive the equations of motion for the normalized momentum-dependent 
density and paring amplitudes  
\begin{equation}
n_{\bk \s} \equiv \frac{\Tr \left( \rho_{0,\a} \cc_{\bk \s} \ca_{\bk \s} \right)}
{\Tr \left( \rho_{0,\a} \right)}
\end{equation} 
and 
\begin{equation}
\D_{\bk } \equiv 
\frac{\Tr \left( \rho_{0,\a} \cc_{\bk \up} \cc_{-\bk \down} \right)}
{\Tr \left( \rho_{0,\a} \right)}.
\end{equation}
From the momentum-dependent observables, the local density and order 
parameter read
\begin{equation}
n = \sum_{\bk \s} n_{\bk \s} \quad{\rm and} \quad \D = \sum_{\bk} \D_{\bk}.
\end{equation}
Notice that, by symmetry, $n_{\bk \up} = n_{-\bk \up} = n_{\bk \down}$ and 
$\D_{\bk} = \D_{-\bk}$.
The equations of motion reads
\begin{equation}
\partial_t n_{\bk\s} = \frac{\Tr \left( {\cal L}_{BCS,\a} \cc_{\bk \s} \ca_{\bk \s} \right)}
{\Tr \left( \rho_{0,\a} \right)} - n_{\bk \s} 
\frac{\Tr \left( {\cal L}_{BCS,\a} \right)}{\Tr \left( \rho_{0,\a} \right)}
\label{eq:dot_nk_normalization}
\end{equation}
and 
\begin{equation}
\partial_t \D_{\bk} = \frac{\Tr \left( {\cal L}_{BCS,\a} \cc_{\bk \up} \cc_{-\bk \down} \right)}
{\Tr \left( \rho_{0,\a} \right)} - \D_{\bk} 
\frac{\Tr \left( {\cal L}_{BCS,\a} \right)}{\Tr \left( \rho_{0,\a} \right)},
\label{eq:dot_dk_normalization}
\end{equation}
where, in both equations, the second term comes 
from the normalization.
In fact, one does not need to compute this normalization 
term  as it is exactly canceled   by the disconnected contractions 
coming from the trace in the first term.
Let us consider the first term in  Eq.~\eqref{eq:dot_nk_normalization}.
Due to the gaussianity of the state 
$\overline{\rho}_{0,\a} \equiv \rho_{0,\a} / \Tr\left( \rho_{0,\a} \right)$, 
the trace is expressed in terms of all the possible contractions 
involving two fermionic lines. 
The contractions divide into two classes: the contractions
that connect one operator of the Liouvillian and one 
operator from $\cc_{\bk \s} \ca_{\bk \s}$ (connected contractions),
and the contractions that connect operators belonging only to the 
Liouvillian or only to $\cc_{\bk \s} \ca_{\bk \s}$ (disconnected contractions).
It is easy to see that the sum of all the terms involving disconnected 
contractions cancels 
with the normalization term in the equations of motion. The same argument holds for Eq.~\eqref{eq:dot_dk_normalization}.
As a result, in order to take into account the normalization,
we eliminate the second term and keep only the connected contractions in 
the first term 
\JoinUp{0.5}{1}{0.5}{1}{nk}
\begin{equation}
\partial_t n_{\bk\s} = 
\Tr \left( \tikzmark{startnk}{{\cal L}_{BCS,\a}[\overline{\rho}_{0,\a}]} 
\tikzmark{endnk}{\cc_{\bk \s} \ca_{\bk \s}} \right)
\end{equation}
and
\JoinUp{0.5}{1}{0.5}{1}{Dk}
\begin{equation}
\partial_t \D_{\bk} = 
\Tr \left( \tikzmark{startDk}{{\cal L}_{BCS,\a}[\overline{\rho}_{0,\a}]} 
\tikzmark{endDk}{\cc_{\bk \up} \cc_{-\bk \down}} \right),
\end{equation}
where the lines indicate contraction connecting the Liouvillian 
and the operator.

The equations of motions can be written in terms of full 
Lindblad and hybrid, \ie $\a <1$, contributions as 
\begin{equation}
\begin{split}
 \partial_t n_{\bk\s} &= 
\left[\partial_t n_{\bk \s} \right]_{\rm Lindblad}
+ (\a-1)
\left[\partial_t n_{\bk \s} \right]_{\rm hybrid} 
\\
\partial_t \D_{\bk\s} &= 
\left[\partial_t \D_{\bk \s} \right]_{\rm Lindblad}
+ (\a-1)
\left[\partial_t \D_{\bk \s} \right]_{\rm hybrid},
\end{split}
\end{equation}
see appendix~\ref{sec:app_eoms} for details.
Upon defining the the complex gap amplitude as
a function  of both the loss and pump rates
\begin{equation}
\Phi \equiv \left[ -|U|+i (\G-P) \right] \D,
\end{equation}
The full Lindbald contributions read
\begin{equation}
\begin{split}
\left[\partial_t n_{\bk \s} \right]_{\rm Lindblad} = 
&- 2 {\rm Im} \left[ \Phi\D_{\bk} \right] - \G n n_{\bk \s} \\
& + 2 P \left( 1- \frac{n}{2} \right)\left( 1 - n_{\bk \s}  \right) 
\end{split}
\label{eq:eom_nk_lindblad}
\end{equation}
and 
\begin{equation}
\begin{split}
\left[ \partial_t \D_{\bk} \right]_{\rm Lindblad}
= & i 2 \epsilon_{\bk} \D_{\bk} - i \Phi (2 n_{\bk \s} - 1 ) - \G n \D_{\bk} \\
& - 2P \left(1 - \frac{n}{2} \right) \D_{\bk} .
\end{split}
\label{eq:eom_deltak_lindblad}
\end{equation}
The hybrid contributions further split into losses and 
pumps terms
\begin{equation}
\begin{split}
\left[\partial_t n_{\bk \s} \right]_{\rm hybrid} & 
=
\left[\partial_t n_{\bk \s} \right]_{\rm hybrid}^{\rm loss}
+ \left[\partial_t n_{\bk \s} \right]_{\rm hybrid}^{\rm pump} \\
\left[\partial_t \D_{\bk} \right]_{\rm hybrid} & 
=
\left[\partial_t \D_{\bk } \right]_{\rm hybrid}^{\rm loss}
+ \left[\partial_t \D_{\bk } \right]_{\rm hybrid}^{\rm pump}
\end{split}
\end{equation}
where the each contribution reads
\begin{equation}
\begin{split}
\left[\partial_t n_{\bk \s} \right]_{\rm hybrid}^{\rm loss} =&
  - \G n \left[   n_{\bk \s}^2 -|\D_{\bk}|^2 \right] +  2 \G \left[ 2 {\rm Re}[\D \D_{\bk}^*] n_{\bk \s} \right],
\end{split}
\end{equation}
\begin{equation}
\begin{split}
\left[\partial_t n_{\bk \s} \right]_{\rm hybrid}^{\rm pump} =&
2 P  \left( 1 -\frac{n}{2}\right) \left[  (1-n_{\bk \s})^2 -|\D_{\bk}|^2 \right] \\
& +  2 P\left[ 2 {\rm Re}[\D \D_{\bk}^*] (1-n_{\bk \s}) \right],
\end{split}
\end{equation}
\begin{equation}
\left[\partial_t \D_{\bk } \right]_{\rm hybrid}^{\rm loss} = 
2 \G \left[ -  n \D_{\bk} n_{\bk \s} + \D n_{\bk \s}^2 - \D^* \D_{\bk}^2
\right],
\end{equation}
and 
\begin{equation}
\begin{split}
\left[\partial_t \D_{\bk } \right]_{\rm hybrid}^{\rm pump} =& 
2 P   \left( 2-n \right) \D_{\bk} (n_{\bk \s} - 1) + \\
& + 2P \left[ \D (1- n_{\bk \s})^2 - \D^* \D_{\bk}^2 \right].
\end{split}
\end{equation}
We notice the loss contributions map into pumps upon the 
 particle-hole transformation $n_{\bk\s} \to 1 -n_{\bk \s}$, 
 and $\D_{\bk} \to - \D_{\bk}^*$.

\section{Results}\label{sec:results}
In this section, we discuss the hybrid-dissipative 
dynamics obtained by solving the equations of motions 
for different quench protocols involving the sudden switch 
of the dissipation rates.
In sections~\ref{res:losses} and~\ref{ref:losse_NHZ} 
we consider only two-body losses, whereas in section~\ref{res:lossespumps}
we consider the simultaneous action of pumps and losses.

\subsection{Depletion dynamics with pair losses}
\label{res:losses}
\begin{figure}
\includegraphics[width=\columnwidth]{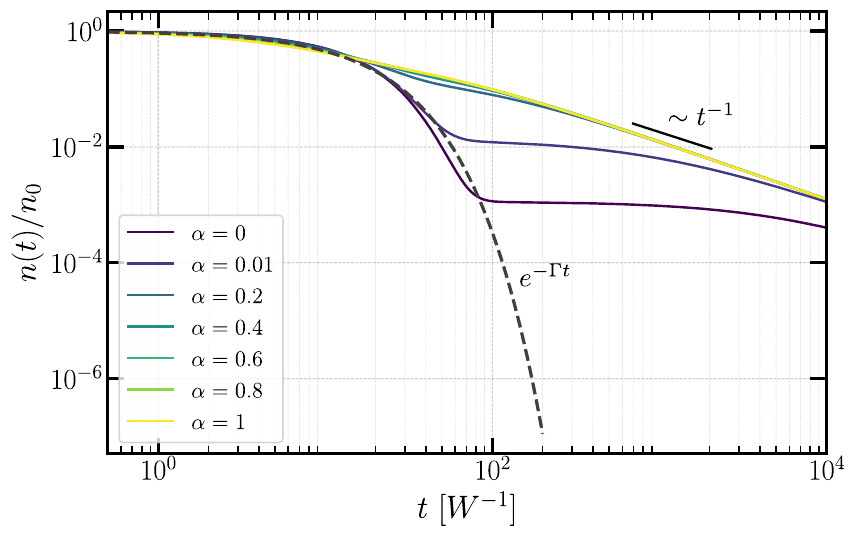}\\
\includegraphics[width=\columnwidth]{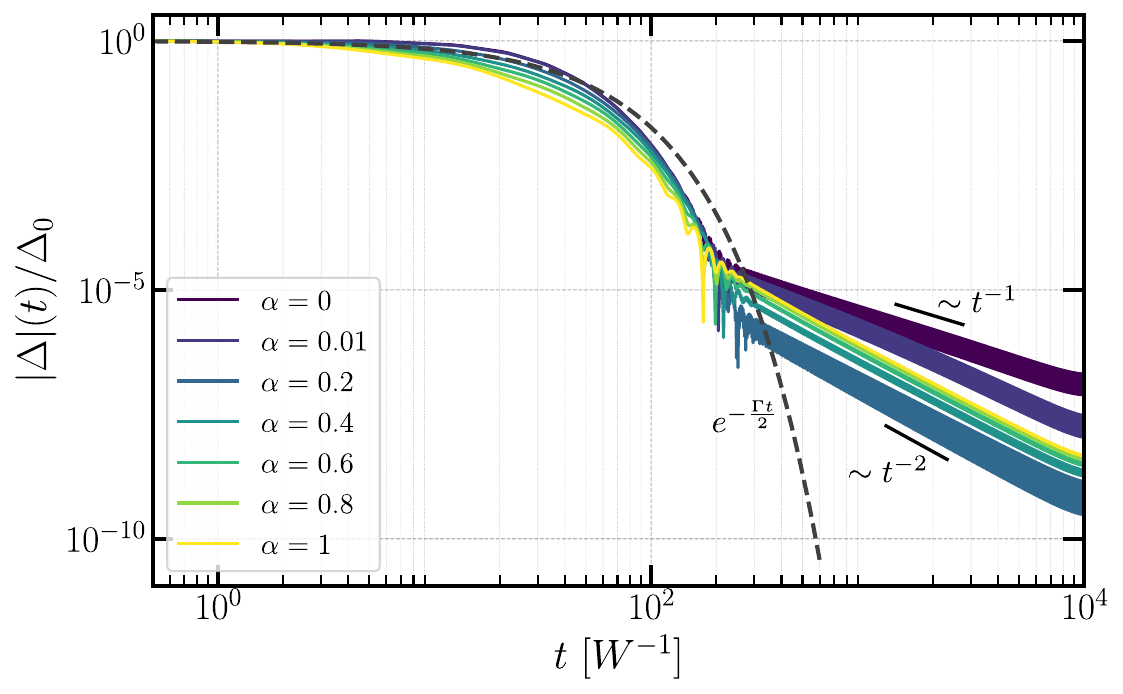}
\caption{Dynamics of the particles density (top panel)
and the superconducting order parameter amplitude (bottom panel) 
for the different values of the parameter $\a$.
$|U|/W=1.0$, $\G/|U| = 0.08$ and $P/|U|=0.0$.}
\label{fig:fig1}
\end{figure}

We consider the sudden switch of the rate of the 
two-body losses $\G(t) = \theta(t) \G$ and set to zero 
the rate of the pumps $P=0$.
Due to the switching the two-particle loss, and in the absence 
of a counterbalancing pump mechanism, the system is expect 
to naturally evolve towards the zero density limit.
In Fig.~\ref{fig:fig1}, we show the depletion dynamics of 
the density and the superconducting order parameter amplitude 
for a fixed values $|U|/W = 1.0$ and $\G/|U|=0.08$ and 
different values of the hybrid parameter $\a$ from the Lindblad 
($\a=1$, yellow lines) to the non-Hermitian ($\a=0$, dark purple lines) limits. 

In the Lindblad limit, the decay of both quantities 
shows an exponential decay at short times, followed 
by universal power-laws $|\D(t)| \sim t^{-2}$ and $n(t) \sim t^{-1}$. 
As discussed in Ref.~\onlinecite{mazza_schiro_PRA_2023},
the power-law decay of the density is characteristic of the 
many-body nature of the jump operators which give rise to 
effective single-particle losses whose rates are proportional 
to the density, see Eqs.~\eqref{eq:eom_nk_lindblad}-\eqref{eq:eom_deltak_lindblad}.
In particular, the rates becomes smaller and smaller as 
the density decreases effectively slowing down the depletion of the system. 
Both dynamics can be understood by looking at the equations 
of motions for density and order parameter amplitude for $\a=1$
\begin{equation}
\partial_t{n}_{\a = 1} = - 4 \G |\D|^2 - \G n^2,
\label{eq:eom_density_lindblad}
\end{equation}
and 
\begin{equation}
\partial_t |\Delta|^2_{\a=1} = - 2 \G |\Delta|^2 + 2 i 
\sum_{\bk \bk'}(\epsilon_{\bk}-\epsilon_{\bk'}) \Delta_{\bk} \Delta_{\bk'}^*.
\label{eq:eom_delta2}
\end{equation}

The equation of motion for the density has a term proportional to the 
order parameter amplitude, which describes the particle depletion due to 
the loss of particles in the condensate, and a term $\sim - \G n^2$ 
which comes from the effective single-particle losses and takes 
into account losses of particles outside the condensate.
At long times, one can neglect the contribution of the 
order parameter, and the equation of motion reduces to $\partial_t n \sim - \G n^2$ 
giving rise to $n(t) \sim \frac{n_0}{(1+\G t)}$.

\begin{figure}
\includegraphics[width=\columnwidth]{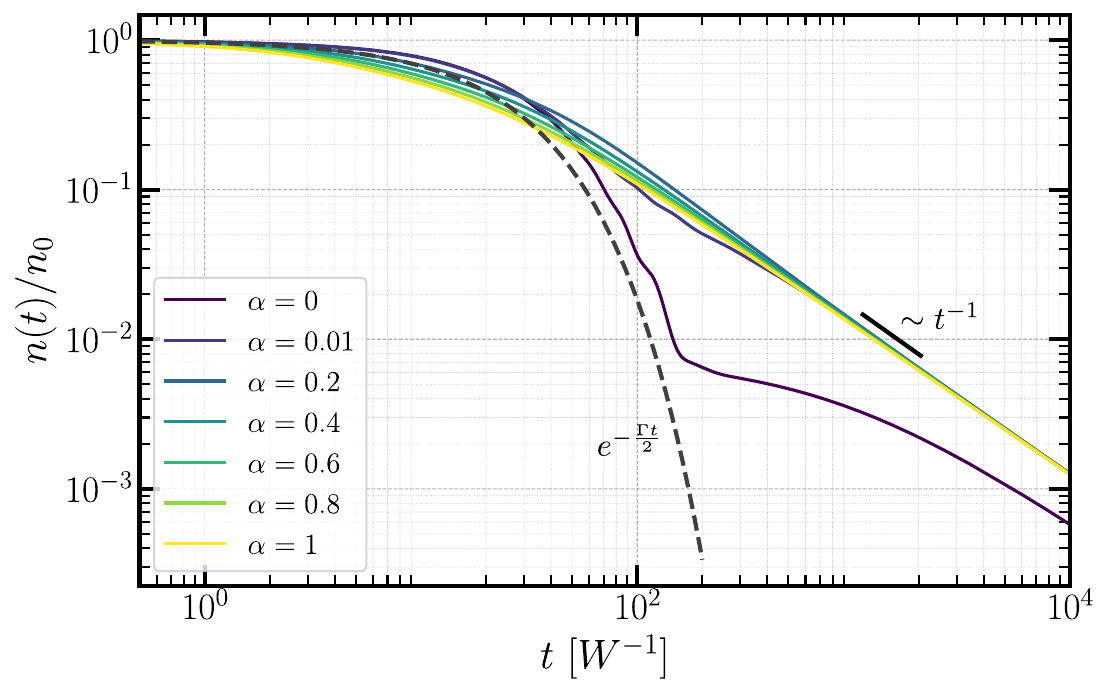}\\
\includegraphics[width=\columnwidth]{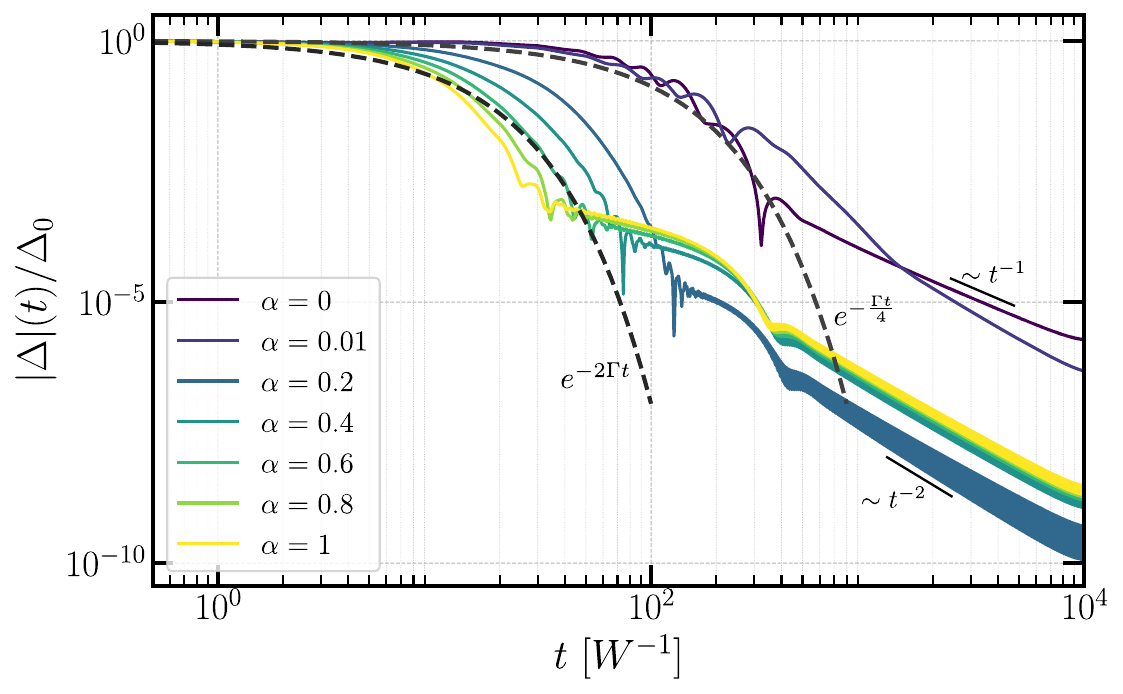}
\caption{Dynamics of the particles density (top panel)
and the superconducting order parameter amplitude (bottom panel) 
for the different values of the parameter $\a$.
$|U|/W=0.5$, $\G/|U| = 0.08$ and $P/|U|=0.0$.}
\label{fig:fig2}
\end{figure}

The dynamics of the order parameter amplitude is 
characterized by the competition between the 
exponential decay and a dephasing term which depends
on the dynamics of the pair amplitudes for each mode.
The exponential decay is the dominant contribution at 
short times. At long times, when $|\Delta|/\Delta_0 \ll 1$, 
the exponential term can be neglected so that the dephasing 
becomes the dominant mechanism $\partial_t |\Delta|^2_{\a=1} \sim 2 i 
\sum_{\bk \bk'}(\epsilon_{\bk}-\epsilon_{\bk'}) \Delta_{\bk} \Delta_{\bk'}^*$.
The $\sim t^{-2}$ decay can be understood by considering the equations of motions for each paring amplitude $\Delta_{\bk}$. Upon neglecting contributions 
from the average pairing amplitude, we obtain $\Delta_{\bk} \sim 
{e^{i 2\epsilon_{\bk} t}}/{(1+\G t)}$, where the 
$(1+\G t)$ denominator comes from the power-law decay of the 
density. This shows that the $\sim t^{-2}$ decay in $\Delta = \sum_{\bk} \Delta_{\bk}$ 
has a $\sim t^{-1}$ contribution coming from the density, and an additional 
$\sim t^{-1}$ contribution from the momentum summation of
the oscillatory terms with phases that accumulate proportionally to $\epsilon_{\bk}$  
for each mode. We also notice that the dephasing term acts 
non-trivially on the exponential decay at short times. 
Indeed, the exponential term alone in the equation~\eqref{eq:eom_delta2} 
would predict a simple $ |\Delta| \sim e^{- \G t}$ decay rate, 
independent on the interaction strength $|U|$.
However, as can be appreciated by comparing Fig.~\ref{fig:fig1} with 
Fig.~\ref{fig:fig2}, showing the depletion 
dynamics for a different value of $|U|/W=0.5$, we find that the short times 
exponential decay evolves as a function of $|U|$. 
 While the interaction strength $|U|$ does not explicitly enter in the 
 exponential decay term of Eq.~\eqref{eq:eom_delta2}, it does 
enters the dynamics of the phase of the pairing amplitude 
 signaling a non-negligible contribution of dephasing also at short times.

Moving away from the Lindblad limit, $\a < 1$, we observe 
a dynamics similar to the Lindblad one occurring on longer   
time scales.
In particular, as $\a$ is decreased, the exponential decay becomes 
slower, and the transition towards the power-law regime gets delayed. 
Physically, this is expected since decreasing $\a$ is interpreted 
as an effective suppression of the dissipative channels due to the 
post-selection of quantum trajectories with reduced number of jumps.
Remarkably, approaching the NH limit, $\a \to 0$, the dynamics sharply changes.

In the purely NH limit, $\a=0$, the density displays 
an exponential decay for an extended interval of time. 
After that, we observe the formation of a quasi-steady state 
characterized by a density plateau. Eventually, 
the dynamics slowly recovers the $t^{-1}$ decay for $tW \gg 1$. 
At the same time, the power-law decay of the order parameter 
suddenly changes from the $\sim t^{-2}$ to a $\sim t^{-1}$ decay.
For infinitesimal $\a$, of the order of $ \a \lesssim 0.01$, 
the dynamics interpolates between the NH and the Lindblad limits.
The same qualitative dynamics, 
with a much less pronounced quasi-steady state density plateau, 
is observed for $|U|/W=0.5$.
By comparing the dynamics of the density and the order parameter for $\a=0$, 
we see that, the NH dynamics, as compared to the Lindblad one, 
is characterized by a faster depletion of particles and a slower 
decay of the order parameter amplitude. 

To get an insight on the above dynamics, we explicitly write the equation of motion for 
the density for arbitrary $\a$
\begin{equation}
\begin{split}
\partial_t{n}_{\a} & = -4 	 \G |\D|^2  - \G n^2  + 2(\a-1) \G \times \\
& \times \sum_{\bk \s}
\left[ \frac{n}{2} \left( |\D_{\bk}|^2-n_{\bk \s}^2 \right) - 2 {\rm Re}(\D \D_{\bk}^*) n_{\bk \s} \right].
\end{split}
\label{eq:eomdensity}
\end{equation}
In equation~\eqref{eq:eomdensity}, the first two terms correspond to the 
Lindbald dynamics, see Eq.~\eqref{eq:eom_density_lindblad}, whereas
the last term is the correction due the hybrid nature of the Liouvillian.
%\deleted{The equation for the order parameter for $\a=0$ is more 
%involved and we present it in the appendix~\ref{something}.}
\begin{figure}
\includegraphics[width=\columnwidth]{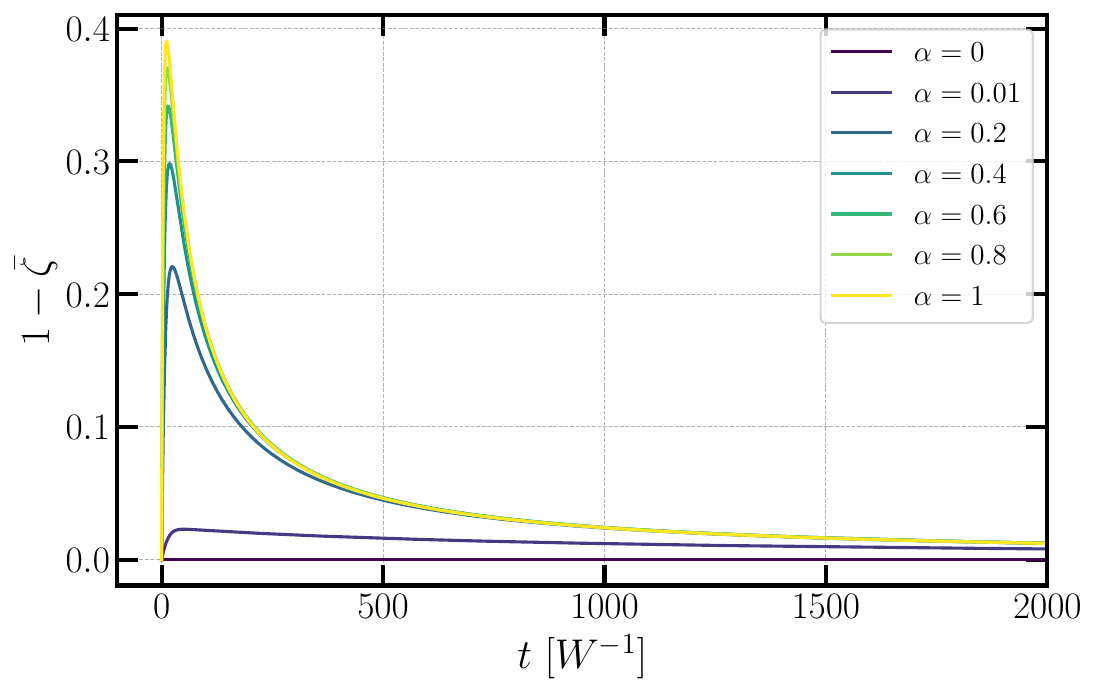}
\caption{Dynamics of the pseudospin length for the different values of 
the parameter $\a$. $U/W=1.0$, $\G/|U| = 0.08$ and $P/|U|=0.0$.}
\label{fig:fig3}
\end{figure}
To make progress, we analyze the equation of motions for the density 
in combination with the dynamics of the pseudospin length, 
defined as 
\begin{equation}
{\zeta}_{\bk} \equiv \sum_{i={x,y,z}} \left[\sigma_{\bk}^i\right]^2
\end{equation}
with $\s_{\bk}^x=2 {\rm Re} \D_{\bk}$, $\s_{\bk}^y=2 {\rm Im} \D_{\bk}$
and $\s_{\bk}^z = 2 n_{\bk \s} -1$. 
%leading to
%\begin{equation}
%{\zeta}_{\bk } = 1 + 4 |\D_{\bk}|^2 - 4 n_{\bk \s}(1-n_{\bk \s}).
%\end{equation}
In the equilibrium BCS state, ${\zeta}_{\bk} = 1$ for each mode. 
Upon switching-on two-particle losses, the pseudo-spin length deviates 
from the equilibrium value, ${\zeta}_{\bk}=1$, and recovers at long times its 
initial value when the system approaches the vacuum state
characterized by $\s_{\bk}^{x} = \s_{\bk}^y =0$, and $\s_{\bk}^z=-1$. 
The evolution of the ${\zeta}_{\bk}$ is governed by the equation
\begin{equation}
\partial_t{\zeta}_{\bk,\a} = -  C_{\bk \a} ({\zeta}_{\bk}-1) - \a\G n n_{\bk \s}
\label{eq:Ldot}
\end{equation}
with $C_{\bk \a}
\equiv -2\G n + (\a-1)\left[ 4 {\rm Re}(\D \D_{\bk}^*) + n (2 n_{\bk\s} + 1) \right]$.
From Eq.~\eqref{eq:Ldot}, we see that, while for $\a=1$ the
pseudo-spin length evolves accordingly to $\partial_t {\zeta}_{\bk} = -2 \G n(\zeta_{\bk}-1)- \G n n_{\bk}$, for $\a=0$ the initial 
pseudo-spin length $\zeta_{\bk}$ becomes  a constant of motion.
The conservation of the pseudo-spin length in the NH limit is shown in Fig.~\ref{fig:fig3}, where we report 
the evolution of the average pseudo-spin length $\overline{\zeta} = \sum_{\bk} \overline{\zeta}_{\bk}$
for different values of $\a$.

The conservation of the pseudo-spin implies that the pairing amplitude 
and the occupation of each mode are constrained by the relation
\begin{equation}
|\D_{\bk}|^2 = n_{\bk \s}\left( 1 - n_{\bk \s} \right).
\label{eq:psL_conservation}
\end{equation}
We can now exploit the conservation of the pseudospin length and use  Eq.~\eqref{eq:psL_conservation} to rewrite the equation of motion for 
the density for $\a = 0$ as 
\begin{equation}
\partial{n}_{\a=0} = - 4 \G |\Delta|^2 - 4 \G n \sum_{\bk} |\Delta_{\bk}|^2 +  
4 \G \sum_{\bk} n_{\bk} {\rm Re}(\D \D_{\bk}^*).
\label{eq:ndot_NH}
\end{equation}
Similar manipulations for the equation of motions of the order parameter 
amplitude leads to 
\begin{equation}
\partial_t |\Delta|^2_{\a=0} \sim - 2 \G |\Delta|^2 + 2 i 
\sum_{\bk \bk'}(\epsilon_{\bk}-\epsilon_{\bk'}) \Delta_{\bk} \Delta_{\bk'}^* + 
8 \G |\D|^4,
\label{eq:eom_delta2_NH}
\end{equation}
where we neglected subleading corrections 
and the quartic term must be understood as an order of magnitude estimate 
coming from the approximation $(\D^*)^2 \sum_{\bk} \D_{\bk}^2 \sim |\D|^2 \sum_{\bk} |\D_{\bk}|^2 \sim |\D|^4$.

Eq.~\eqref{eq:eom_delta2_NH} reveals that the dynamics of the order parameter 
amplitude is essentially the same of the Lindblad case, with an exponential decay 
and a dephasing term. The only difference is in the $\sim 8 \G |\D|^4$ term 
which partially counterbalances the exponential term and is likely to be 
responsible of the slower decay of $|\D|$ upon decreasing $\a$. 
The power-law decay at long-times has the same dephasing origin as in the 
Lindblad case. However, the exponent changes from $t^{-2}$ to $t^{-1}$ due 
to the fact that the additional $\sim t^{-1}$ contribution coming from the 
power-law decay of the density is now suppressed.\

Considering the equation of motion for the density, we observe that, as a 
consequence of the pseudo-spin conservation, the $-\G n^2$ terms that was 
describing effective single particle losses and was responsible of the power law decay 
of the density for $\a > 0$, disappears in the NH limit. This means that, effective 
single particle losses are suppressed in the NH limit, and the depletion of particle 
is completely dragged by the decay of the order parameter. 
At short times, the last two terms in Eq.~\eqref{eq:ndot_NH} 
approximately cancels each other, so that we can approximate $\partial_t{n}_{\a=0} \sim - 4 \G |\D|^2$. 
By assuming a simple exponential decay with rate $\g$ (which depends itself on the dissipation rate $\G$) 
for the order parameter $|\D| \sim |\D_0| e^{- \g t} $, we obtain
$n_{\a=0} \sim n_0 +\frac{2 \G}{ \g} |\D_0|^2 \left( e^{-2 \g t} -1 \right)$
which implies that, eventually, a plateau is expected for $\g t \gg 1$.
The value of the plateau density depends on the actual dynamics of 
the order parameter amplitude that, as can be appreciated from the 
reference exponential decays in the figures, is richer than the simple 
decay used in this argument. Nonetheless, we observe that the above 
argument is further supported by the fact that, in the numerics, 
the exponential decay rate of the density is found to be 
roughly twice the decay rate of the order parameter amplitude. 

At long times, we can assume that, due to dephasing,
$|\sum_{\bk} \D_{\bk}| \ll \sum_{\bk} |\Delta_{\bk}|$ and retain 
only the second term in Eq.~\eqref{eq:ndot_NH}. 
This leads to 
$\partial_t {n}_{\a=0} \sim - 4 \G n \sum_{\bk} |\D_{\bk}|^2 = - \G n \sum_{\bk} 2 n_{\bk} - n_{\bk}^2$ 
where, in the last step, we used again the pseudo-spin length conservation.
For very long-times $n_{\bk}^2 \ll n_{\bk}$ and the dynamics reduces 
to $\partial_t {n}_{\a=0} \sim - 2 \G n^2$ which 
justifies the $\sim t^{-1}$ decay of the density for $tW \gg 1$.
Despite the apparent similarity between such a long-time 
power-law decay of the density for $\a=0$ and the 
same power-law decay for $\a=1$, we stress that the 
two behaviors have completely different origin.
In the latter case, $\a=1$, the power-law decay is due to 
the effective single-particle losses and it 
characterizes the dynamics at all times. 
On the contrary, for $\a=0$, effective single-particle losses 
are suppressed and the power-law decay emerges only at long 
times when the dephasing decay of the order parameter becomes 
the dominant contribution. 
%namely when $|\sum_{\bk} \D_{\bk}|^2$ 
%can be neglected with respect to $\sum_{\bk} |\D_{\bk}|^2$.

\begin{figure}
\includegraphics[width=0.99\columnwidth]{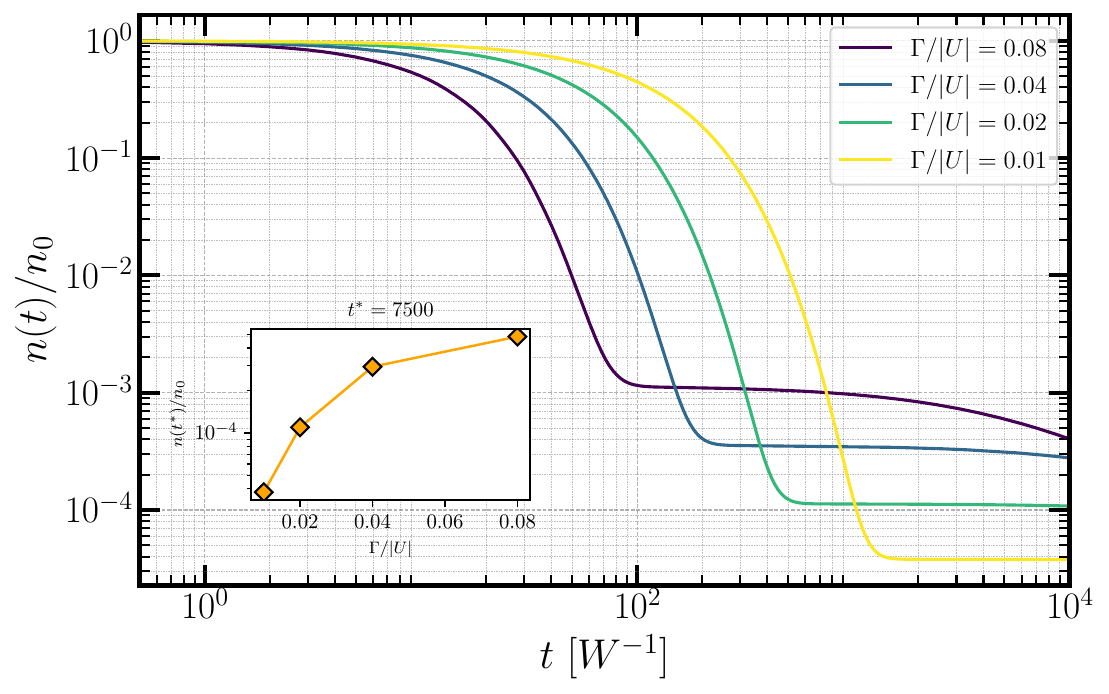}
\includegraphics[width=0.99\columnwidth]{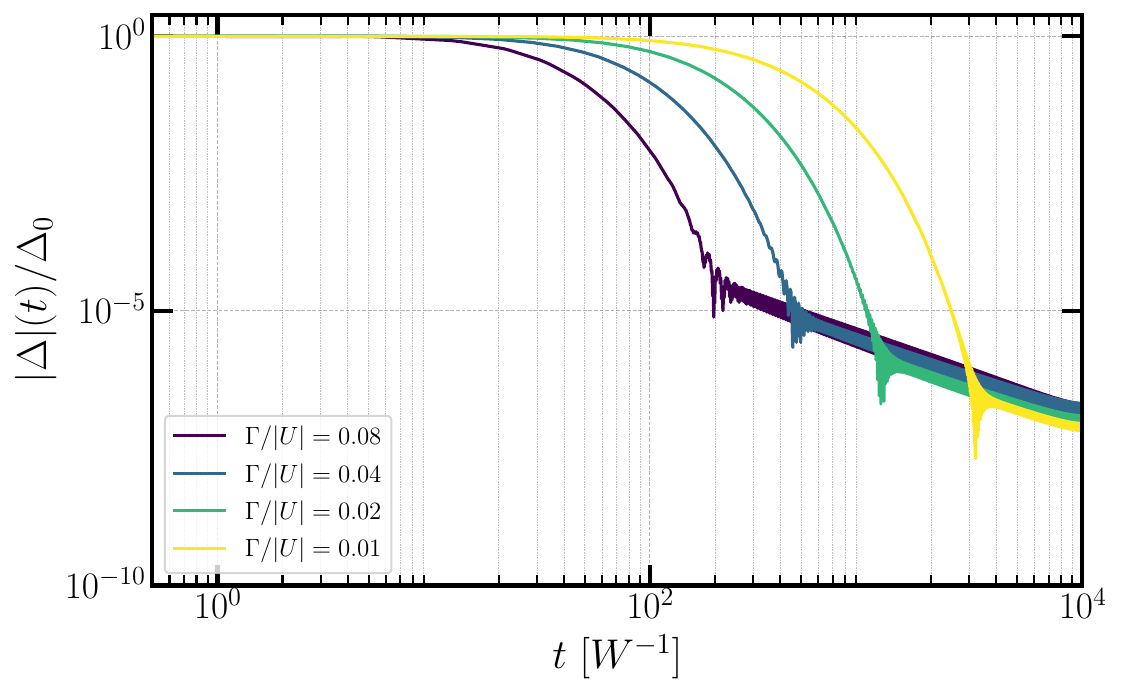}
\caption{Dynamics of the particle density (top panel) 
and the superconducting order parameter amplitude 
(bottom panel) in the NH limit for different values 
of the dissipation rate $\G$.
$|U|/W=1.0$, $\a=0$ and $P/|U|=0.0$.
Inset in the top panel: density in the 
quasi-steady plateau as a function of 
the dissipation rate.}
\label{fig:fig4}
\end{figure}

\subsection{Non-Hermitian Zeno Effect}
\label{ref:losse_NHZ}
We now discuss the direct consequences of the dynamics 
in the NH limit described in the previous section.
In the NH limit, the pseudospin length conservation 
suppresses effective single particle losses and 
the decay of density is completely tied to the decay 
of the order parameter. 
This leads to an exponential decay at short times 
ending up in a plateau which slowly recovers the 
power-law decay $\sim t^{-1}$ at later times 
due to the dephasing of the order parameter.
This behaviour has interesting consequences. 
First of all, the onset of the $\sim t^{-1}$ 
decay is expected to occur earlier the faster 
the decay of the order parameter.
Moreover, the faster the decay of the order parameter, 
the smaller the number of particles that can be depleted  
before the plateau set in. 
In other words, as already observed in Figs.~\ref{fig:fig1}-\ref{fig:fig2}, 
the slower the decay of the order parameter the more effective 
the depletion of particles due to NH two-body losses.
These observations point towards a Non-Hermitian Zeno (NHZ) 
effect~\cite{Syassen2008,Garcia-Ripoll2009,Froml2019,Biella2021,Rossini2021,Scarlatella2020} in which, before the power-law decay due to dephasing sets in, 
the density plateau in the quasi-steady state increases 
as a function of $\G$. 
In Fig.~\ref{fig:fig4}, we report evidence of such a NHZ 
effect by showing the NH dynamics of the density and 
superconducting  order parameter amplitude for several values of the 
dissipation rate $\Gamma.$ 
As it is clear from the inset, the plateau density 
increases as function of the decay rate. 
Analogously, we observe that the smaller the $\G$
the smaller the amplitude of the order parameter 
reached after the exponential decay.

\subsection{Driven-dissipative dynamics with pair losses and pumps}
\label{res:lossespumps}

\begin{figure}
\includegraphics[width=\columnwidth]{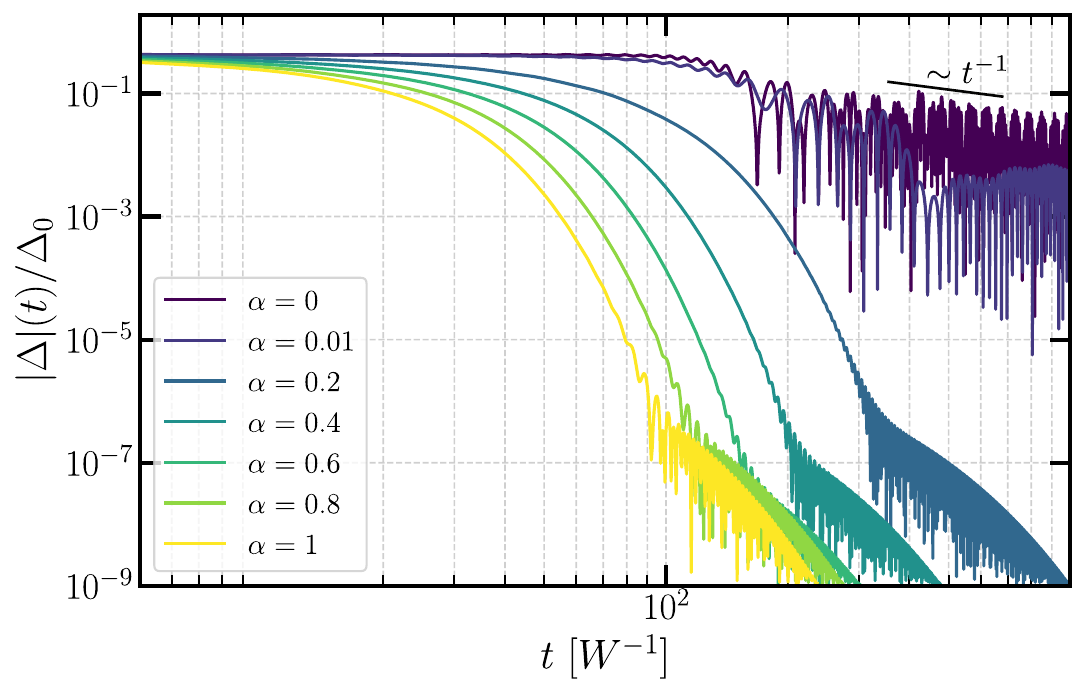}
\caption{Dynamics of the order parameter with simultaneous pair pumps and losses
for different values of the parameter $\a$.
$|U|/W=1.0$ and $\G/|U| = P/|U|= 0.08$.}
\label{fig:fig5}
\end{figure}

We now consider the driven-dissipative case in which 
both two-body losses and pumps are simultaneously switched-on. 
We consider the equally balanced case $\G = P$ which preserves 
particle-hole symmetry so that the density remains constant 
in time $n(t) =1$.
Due to the homogeneous rates of pairs injection and pairs removal, 
the system is expected to continuously absorb energy and the order 
parameter is expected to decay until an infinite temperature state 
is eventually reached.
In Fig.~\ref{fig:fig5}, we show the dynamics of the order parameter 
for fixed value of the dissipation rate and various hybrid 
parameters $\a$ from the Lindbald, $\a=1$, to the NH, $\a=0$, limits. 
Starting from $\a=1$, we see that in the 
driven-dissipative case the decay of the order parameter 
has always an exponential character.
This is expected as, in contrast to the previous case, 
the dissipation rates for the effective single-particle dissipators, 
namely $n\G$ for the losses and $(2-n)P$ for the pumps, remain 
constant in time.
As $\a$ is decreased, the exponential decay becomes slower, 
as expected by the effective reduction of the effects of 
the jump operator. 
The dynamics smoothly evolves as a function of $\a$ 
until, similarly to what discussed for the only 
loss case, it sharply changes approaching the NH limit. 
For $\a = 0$, the order parameter amplitude remains more ore less 
constant for an extended interval of time until a rapidly 
oscillating power-law decay $\sim t^{-1}$ sets 
in for $tW \gtrsim 100$.

\begin{figure}
\includegraphics[width=\columnwidth]{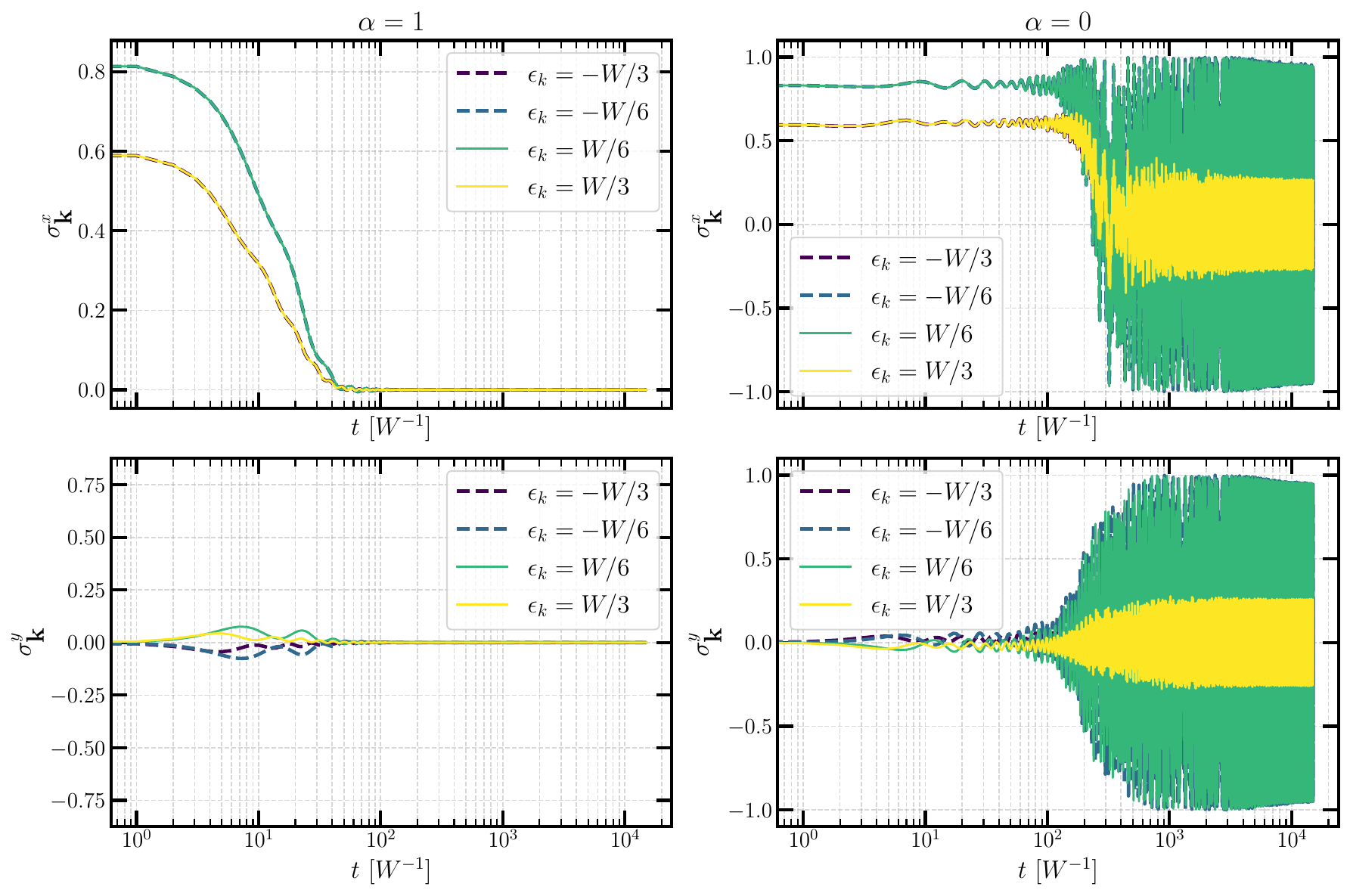}
\caption{Dynamics of the pairing amplitudes $\sigma_{\bk}^x$ 
(top panels) and $\sigma_{\bk}^y$ (bottom panels) for $\a =1$ (left panels)
and $\a=0$ (right panels) for selected modes 
above and below the Fermi level.
$|U|/W=1.0$ and $\G/|U| = P/|U|= 0.08$.}
\label{fig:fig6}
\end{figure}

\begin{figure}
\includegraphics[width=\columnwidth]{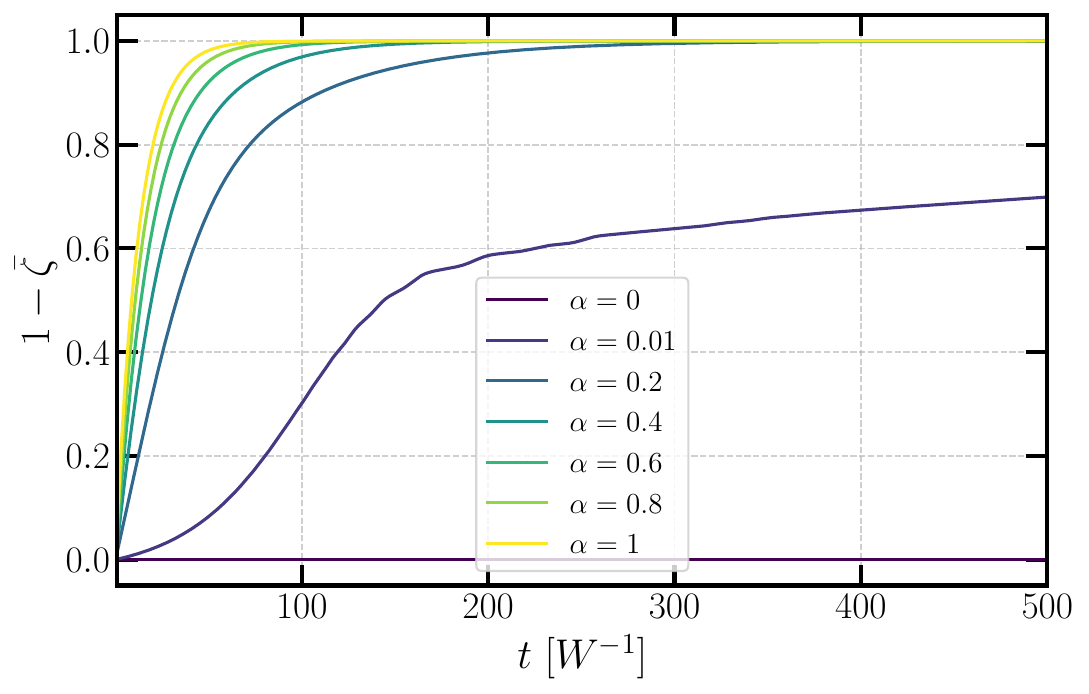}
\caption{Dynamics of the average pseudo-spin length 
for different values of $\a$. $|U|/W=1.0$ and $\G/|U| = P/|U|= 0.08$.}
\label{fig:fig7}
\end{figure}

The power-law decay of the order parameter suggests that, in the NH limit, 
the dynamics is governed by a purely dephasing mechanism.
In contrast, in the Lindblad limit, the exponential decay 
indicates the decay of the pairing amplitude of each mode.
This is clearly seen in Fig.~\ref{fig:fig6}, 
where we report the dynamics of $\s_{\bk}^{x}$ 
and $\sigma_{\bk}^y$ for representative modes 
at selected energies above and below the Fermi energy.
We observe that, for $\a=1$, the pairing amplitudes of 
the single modes all decay to zero. In contrast, for $\a=0$, 
these quantities oscillate with constant amplitude, thus 
showing that the fast oscillating power-law decay of Fig.~\ref{fig:fig5} 
is completely governed by the destructive interference among the different 
modes.

To understand the origin of such a sharp difference between the 
dynamics in the Lindblad and NH limits, we resort once again to the 
dynamics of the pseudo-spin length $\zeta_{\bk}$.
Analogously to the case of only losses, the NH 
dynamics preserves the pseudo-spin length, see Fig.~\ref{fig:fig7}. 
It is straightforward to see that the 
conservation of the pseudo-spin length hinders the decay to zero of  
the pairing amplitudes of the single modes. 
Indeed, if $\s_{\bk}^x = \s_{\bk}^y = 0$  the 
pseudo-spin conservation would require $\s_{\bk}^z = \pm 1$.
Such a pseudo-spin configuration  can be realized only 
for the completely filled or completely empty states 
that, due to the perfect balance between pair pumps and losses, 
can never be reached.
On the contrary, due to the unconstrained evolution of the pseudo-spin 
length, the $\s_{\bk}^x = \s_{\bk}^x = 0$  state can always 
be reached within the Lindblad dynamics.

\begin{figure}
\includegraphics[width=\columnwidth]{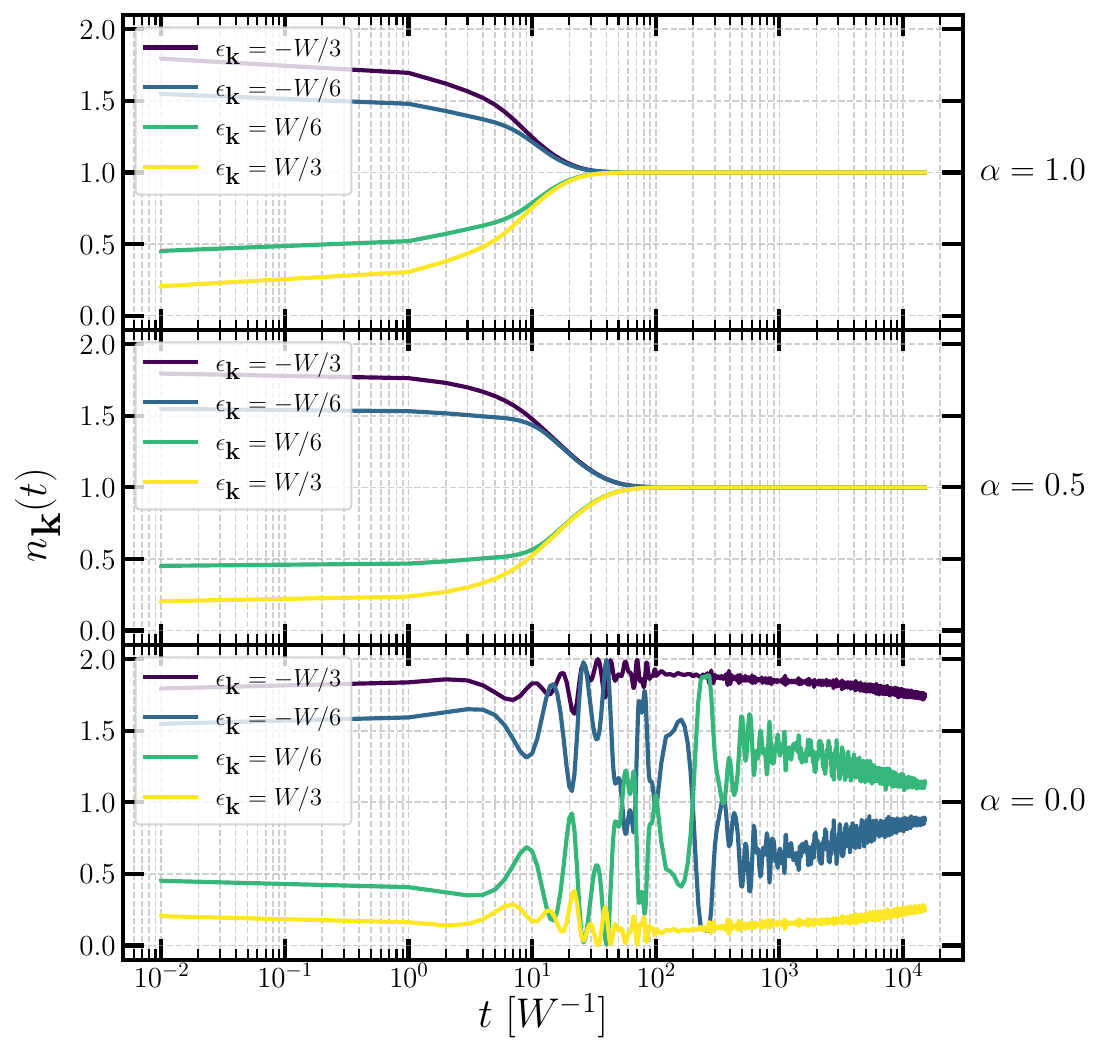}
\caption{Dynamics of the occupation numbers 
for selected modes above and below the Fermi level
for $\a=1.0$ (top), $\a=0.5$ (middle) and $\a=0.0$ (bottom).
$|U|/W=1.0$ and $\G/|U| = P/|U|= 0.08$.}
\label{fig:fig8}
\end{figure}

The above observations show that the Lindblad and the NH 
driven-dissipative dynamics are characterized by the distinct 
steady states. To fully characterize the steady states reached 
by the different non-unitary dynamics, we plot in Fig.~\ref{fig:fig8}
the occupation number $n_{\bk\s} = \frac{1+\s_{\bk}^z}{2}$
for selected modes above and before the Fermi level.
In the Lindblad limit, the dynamics converges towards 
the infinite temperature state with homogeneous occupation 
for all modes $n_{\bk \s} = \frac{1}{2}$ or, equivalently, $\s_{\bk}^z = 0$. 
The same occurs for the hybrid dynamics with $\a<1$.
On the contrary, in the NH limit, we observe that 
the occupations of the modes are characterized by a 
strongly oscillating dynamics. 
Interestingly, after a certain time, the strong 
oscillations leads to a population inversion
with states below the Fermi energy becoming 
less populated than the states above. 
This shows how in the NH limit the system skips the 
infinite temperature steady states and evolves towards a 
highly non-equilibrium steady state akin to a negative 
temperature state. 

\section{Conclusions}\label{sec:conclusions}
We discussed the non-unitary dynamics of dissipative fermions 
described by a hybrid Lindblad evolution, interpolating between 
the full unconditional Lindblad dynamics and the full post-selection 
of quantum jump trajectories captured by the no-click NH dynamics. 
By tuning the control parameter $\alpha$, between $\alpha=0$ (NH dynamics) 
and $\alpha=1$ (Lindblad) we investigate the role of quantum jumps on 
the dynamics of the density matrix. 
To tackle the problem, we have introduced time-dependent 
variational principles based on actions defined 
for arbitrary values of $\a$. The variational principles 
naturally take into account the state normalization in the hybrid 
Lindbladian evolution and, in the NH limit $\a=0$, reduce 
to an extended Dirack-Frenkel for NHH.

We have applied the variational principles to the 
dynamics of driven-dissipative BCS superconductors 
described by an attractive Hubbard model 
in the presence of two-body losses, and 
in the simultaneous presence of both, equally balanced, 
two-body losses and pumps.
In the first case, the non-unitary dynamics leads 
to the complete depletion of the system, whereas 
in the driven-dissipative case the non-unitary 
dynamics with perfectly balanced pumps and losses 
leads to the formation of non-trivial steady-states.

By considering the above non equilibrium protocols,
we explored the evolution of the 
non-unitary dynamics from the Lindblad, $\a=1$, to the NH limits, $\a=0.$
In all the discussed cases, we showed how 
the quantum jumps play a crucial for the non-unitary dynamics. 
Starting from $\a=1$, we progressively reduce the relative 
weight of the quantum jumps, $\a<1$, and show how the 
non-unitary dynamics is equivalent to a 
Lindblad dynamics occurring on longer time scales.
This suggests that the suppression of the relative 
weight of the quantum jumps can be understood 
as an effective reduction of the dissipation rate.
In contrast, in the no-click limit, $\a=0$, when the quantum 
jumps are completely neglected, we observe sharp qualitative 
changes in the dynamics for what concerns both the evolution 
towards the steady state, and the properties of the 
steady state. 
We show that such a sharp qualitative changes in 
the non-unitary dynamics are due to the an exact 
conservation of the pseudospin length occurring for $\a=0$.
In the case of the depletion dynamics, the pseudospin length 
conservation determines the suppression of the effective 
single-particle losses so that the decay of the 
density becomes completely tied to the decay of 
the order parameter. 
For the driven-dissipative case, the same conservation
hinders the exponential decay of the order parameter 
leading to a purely dephasing dynamics characterized 
by a slow power-law of the order parameter.

By analysing the dynamics in different regimes,
we showed that the NH dynamics leads to 
non-trivial steady states that cannot be reached 
in the presence of quantum jumps.
For the depletion dynamics, our results 
show that the evolution towards the vacuum state 
occurs through a quasi-steady state characterized 
by a density plateau.
The particle suppression is the more effective the faster the 
decay of the order parameter, leading to the emergence of a 
Non-Hermitian Zeno effect characterized by quasi-steady states 
with density increasing as a function of the dissipation rate.
For the driven-dissipative case, our results show that 
the the NH dynamics drives the systems 
towards the a highly non-equilibrium steady state 
with population inversion.

\begin{acknowledgments}
G.M.~acknowledges support by the MUR Italian 
Minister of Research through a "Rita-Levi Montalcini"
fellowship. M.S. acknowledges financial support
from the ERC consolidator Grant No. 101002955 - CONQUER.
\end{acknowledgments}

\bibliography{refs_bcs_diss}

\onecolumngrid

\appendix

\section{Hybrid DMVP with Gaussian states}
\label{app:DMVP_Guassian}
In this section we report details of the 
application of the DMVP to Gaussian 
states discussed in Sec.~\ref{sec:non_unitary_TDVP}. 
We start from the case $\a=1$. 
For this, we need to compute the trace
\begin{equation}
\Tr \left(\raux {\cal L}_{1}[\rho_0]  \right) = 
- i  \Tr \left( \raux \left[ {\cal H},\rho_0 \right] \right) + 
\Tr \left( \raux \d{\cal L}_1[\rho_0] \right)
\label{eq:trace_general_gaussian}
\end{equation}
with 
\begin{equation}
{\cal H} = {\cal H}_0 + {\cal H}_{int} = \sum_{\a \b} t_{\a \b} \cc_\a \ca_{\b}
+ \frac{1}{2}
\sum_{\a \b \g \d} U_{\a \b \g \d} 
\cc_\a \cc_{\b} \ca_{\g} \ca_{\d}
\end{equation}
and 
\begin{equation}
\d{\cal L}_1[\rho_0] = \sum_{ab cd}
\G_{abcd}
\left[
\ca_{c} \ca_{d} \rho_0 \cc_{a} \cc_{b}
-
\frac{1}{2}
\left\lbrace 
\cc_{a} \cc_{b} \ca_{c} \ca_{d} , \rho_0
\right\rbrace 
\right]
\end{equation}
where 
we defined $\G_{abcd} \equiv \sqrt{\k_{ba} \k_{cd}}.$
Due to the gaussian nature of the state $\rho_0$, the 
goal is to bring the trace Eq.~\eqref{eq:trace_general_gaussian} 
in the form 
\begin{equation}
\Tr \left(\raux {\cal L}_{1}[\rho_0]  \right) = 
\Tr \left(\raux {\cal L}_{HF}[\rho_0]  \right) 
\end{equation}
where ${\cal L}_{HF}$ is an effective Hartree-Fock 
Liouvillian containing only single particle terms.
We see that the unitary term involving the non-interacting 
Hamiltonian ${\cal H}_0$ is already in this form so that we have 
to compute only the terms involving the interacting part of the  
Hamiltonian ${\cal H}_{int}$ and the many-body dissipator. 
Let us start from the interacting part of the Hamiltonian
\begin{equation}
\Tr \left( \raux \left[{\cal H}_{int}, \rho_0\right] \right) = 
\sum_{abcd} U_{abcd} 
\Tr \left( \raux \left[\cc_{a} \cc_{b} \ca_{c} \ca_{d}, \rho_0\right] \right)
\label{eq:tr_Hint}
\end{equation} 
We now express the right-hand side of Eq.~\eqref{eq:tr_Hint} as 
a trace over the gaussian state $\rho_0$
\begin{equation}
\Tr \left( \raux \left[{\cal H}_{int}, \rho_0\right] \right) = 
\sum_{abcd} U_{abcd} 
\Tr \left( \rho_0  \raux \cc_{a} \cc_{b} \ca_{c} \ca_{d}  \right)
-
U_{abcd} 
\Tr \left( \rho_0  \cc_{a} \cc_{b} \ca_{c} \ca_{d} \raux \right).
\end{equation}
We now can compute the trace by taking all the contractions 
involving two fermionic lines. 
For the sake of this example we will assume, for simplicity, that 
the anomalous contractions are zero, namely 
$\Tr \left(\rho_0 \cc_{a} \cc_{b} \right) 
= \Tr \left(\rho_0 \ca_{a} \ca_{b} \right) = 0.$ 
For example, the first term in the RHS of the previous equation reads
\JoinUp{0.5}{1}{0.5}{1}{1}
\JoinUp{0.5}{1}{0.5}{1}{2}
\JoinUp{0.5}{1}{0.5}{1}{3}
\JoinUp{0.5}{1}{0.5}{1}{4}
\begin{equation}
\Tr \left( \rho_0  \raux \cc_{a} \cc_{b} \ca_{c} \ca_{d}  \right)
 = 
 \Tr \left( \rho_0  \raux 
 \tikzmark{start1}{\cc_{a}} \cc_{b} \ca_{c} 
 \tikzmark{end1}{\ca_{d}} \right) 
 + 
  \Tr \left( \rho_0  \raux 
 {\cc_{a}} 
 \tikzmark{start2}{\cc_{b}} 
 \tikzmark{end2}{\ca_{c}}
 {\ca_{d}} \right) 
 -
  \Tr \left( \rho_0  \raux 
 \tikzmark{start3}{\cc_{a}} 
 {\cc_{b}} 
 \tikzmark{end3}{\ca_{c}}
 {\ca_{d}} \right)  
  -
  \Tr \left( \rho_0  \raux 
 {\cc_{a}} 
 \tikzmark{start4}{\cc_{b}} 
 {\ca_{c}}
 \tikzmark{end4}{\ca_{d}} \right)  
\end{equation}
where,  in each term, the sign in each term 
takes into account the parity of the permutation.

Each permutation can be written as 
\begin{equation}
\Tr \left( \rho_0  \raux \tikzmark{starta}{\cc_{a}} \cc_{b} \ca_{c} \tikzmark{enda}{\ca_{d}}  \right)
 = \Tr \left( \rho_0 \cc_{a} \ca_{d} \right) \sum_{\rm contractions} 
 \Tr \left( \rho_0  \tikzmark{startb}{\raux}  \tikzmark{endb}{ \cc_{b} \ca_{c}}  \right)
\end{equation}
where the summation over contractions on the RHS indicates the sum
over all the possible contractions involving the extraction of two 
fermionic lines from the operator $\raux \cc_{b} \ca_{c}$.
It is straightforward to see that this summation exactly 
reconstructs the trace of the operator over $\rho_0$, namely
\JoinUp{0.5}{1}{0.5}{1}{a}
\JoinUp{0.5}{1}{0.5}{1}{b}

\begin{equation}
\Tr \left( \rho_0  \raux \tikzmark{startc}{\cc_{a}} \cc_{b} \ca_{c} \tikzmark{endc}{\ca_{d}}  \right)
 = \Tr \left( \rho_0 \cc_{a} \ca_{d} \right) \Tr \left( \rho_0  {\raux}  { \cc_{b} \ca_{c}}  \right).
\end{equation}
\JoinUp{0.5}{1}{0.5}{1}{c}
By iterating the above procedure for all the terms, the first term in Eq.~\eqref{eq:trace_general_gaussian} reads
\begin{equation}
\Tr \left( \raux \left[{\cal H}, \rho_0\right] \right) =
\Tr \left( \raux \left[{\cal H}_{HF}[\rho_0], \rho_0\right] \right)
\end{equation}
with 
\begin{equation}
{\cal H}_{HF} = 
\sum_{ab} t_{ab} \cc_{a} \ca_{b}
+ \frac{1}{2}\sum_{abcd} (U_{abcd}-U_{abdc}) 
\left[
\Tr\left( \rho_0 \cc_{a} \ca_{d} \right) \cc_{b} \ca_{c} 
+ \Tr\left( \rho_0 \cc_{b} \ca_{c} \right) \cc_{a} \ca_{d} 
\right].
\end{equation}
By using the properties of the interaction matrix elements 
$U_{abcd} = U_{b a d c}$ we get
\begin{equation}
{\cal H}_{HF} = 
\sum_{ab} t_{ab} \cc_{a} \ca_{b}
+ \sum_{abcd} (U_{abcd}-U_{abdc}) 
\Tr\left( \rho_0 \cc_{a} \ca_{d} \right) \cc_{b} \ca_{c}.
\end{equation}
By using the same contraction scheme to dissipator term in 
Eq.~\eqref{eq:trace_general_gaussian}, we obtain
\begin{equation}
\Tr \left( \raux \d{\cal L}_1[\rho_0] \right) = 
\Tr \left( \raux \d{\cal L}_{HF}[\rho_0]  \right)
\end{equation}
with 
\begin{equation}
\d {\cal L}_{HF} [\rho_0]
= 
\sum_{abcd}
\left( \G_{abcd} - \G_{abdc} \right)
\left( 
\Tr \left(\rho_0 \cc_{a} \ca_{d} \right)
\left[ \ca_{c} \rho_0 \cc_{b} - \frac{1}{2} \left\lbrace \cc_{b} \ca_{c},\rho_0 \right\rbrace \right] + \left[ (a \leftrightarrow b)\phantom{\frac{1}{2}} (c\leftrightarrow d )\right]\right)
\end{equation}
where the second term is obtained from the first one 
upon exchanging $(a \leftrightarrow b)$ and $(c \leftrightarrow d)$.
The dissipator couplings $\G_{abcd}$ satisfy the same index permutation property
of the interaction terms $\G_{a b cd} = \G_{ba dc} $, so that 
\begin{equation}
\d {\cal L}_{HF} [\rho_0]
= 
2 \sum_{abcd}
\overline{\G}_{abcd}
\Tr \left(\rho_0 \cc_{a} \ca_{d} \right)
\left[
\ca_{c} \rho_0 \cc_{b} - \frac{1}{2} \left\lbrace \cc_{b} \ca_{c},\rho_0 \right\rbrace
\right],
\end{equation}
where we defined $\overline{\G}_{abcd} \equiv \G_{abcd} - \G_{abdc}.$
Eventually, the action reads
\begin{equation}
{\cal S} = \int dt \Tr \left[\raux \left( \partial_t \rho_0 -  {\cal L}_1[\rho_0] \right) \right] 
= 
\int dt \Tr \left[\raux \left( \partial_t - \rho_0 {\cal L}_{HF}[\rho_0] \right) \right]
\end{equation}
and by taking the variation with respect to $\raux$ we obtain that 
the variational density matrix evolves with the 
effective single particle Liouvillian
\begin{equation}i \partial_t \rho_0(t) = -i \left[ {\cal H}_{HF}[\rho_0],\rho_0(t) \right] + \d{\cal L}_{HF}[\rho_0(t)].
\end{equation}
Following the same procedure for normalized states, the $\a < 1$ case 
is straightforwardly obtained, leading to 
\begin{equation}
\partial_t \rho_{0,\a}(t)
= - i 
\left[ {\cal H}_{HF}[\rho_{0,\a}],\rho_{0,\a}\right]
+ \a \d {\cal L}_{HF}[\rho_{0,\a}] + (\a-1) \sum_{a b c d} \overline{\G}_{abcd}
\Tr \left(\rho_{0,\a} (t) \cc_{a} \ca_{d} \right)
\left\lbrace \cc_{b} \ca_{c},\rho_{0,\a} \right\rbrace.
\end{equation}

\section{Hybrid DMVP for superconductivity with many-body dissipation}
\label{sec:hybrid_DMVP_pumps}
In this section, we discuss the detailed derivation of the hybrid BCS Liouvillian 
introduced in Sec.~\ref{sec:bcs_hybrid_liouvillian} of the main text.
We consider only the case of two-body pumps as the loss-case is derived 
by particle-hole transformation.
In order to apply the hybrid DMVP we compute the 
action by calculating the trace of the Liouvillian over the auxiliary state $\raux$
\begin{equation}
\frac{\Tr \left( \raux {\cal L}_{\a}[\rho_{0,\a}] \right)}
{\Tr \left( \rho_{0,\a} \right)}
=
\Tr \left( \raux {\cal L}_{\a}[\overline{\rho}_{0,\a}] \right)
= 
- i \Tr \left( \raux \left[ {\cal H}, \overline{\rho}_{0,\a} \right] \right) 
+ \Tr \left( \raux \d {\cal L}_{\a}[\overline{\rho}_{0,\a}] \right),
\label{eq:liouvillian_trace_pumps}
\end{equation}
with hybrid dissipator $\d {\cal L}_{\a}$ defined by the 
jump operators $L_i = \sqrt{2 P} \cc_{i \up} \cc_{i \down}.$
In Eq.~\eqref{eq:liouvillian_trace_pumps} we defined the normalized 
variational density matrix as
\begin{equation}
\overline{\rho}_{0,\a} \equiv \frac{\rho_{0,\a}}{\Tr \left( \rho_{0,\a} \right)}.
\end{equation}
As seen in the previous section, the first trace 
leads to the Hartree-Fock decoupling of the 
unitary Hamiltonian which, by inclusion of the anomalous contraction terms,
is represented by the standard BCS unitary Liouvillian
\begin{equation}
\Tr \left( \raux \left[ {\cal H}, \overline{\rho}_{0,\a} \right] \right)  = 
\Tr \left( \raux \left[ H_{BCS,0}, \overline{\rho}_{0,\a} \right] \right) 
\end{equation}
where, upon defining the superconducting order 
parameter $\D(t)\equiv \Tr \left( \overline{\rho}_{0,\a} \cc_{i\up} \cc_{i \down} \right)$,
\begin{equation}
H_{BCS,0} = \sum_{ij \s} t_{ij} \cc_{i\a} \ca_{j\s} 
-|U|\D(t) \sum_{i} \ca_{i \down} \ca_{i\up} +{\rm h.c.}.
\end{equation}
We now focus on the trace of the dissipator
\begin{equation}
\Tr \left( \raux \d {\cal L}_{\a}[\overline{\rho}_{0,\a}] \right) = 
P \sum_i 
2 \a \Tr \left( \raux \pairc \overline{\rho}_{0,\a} \paira \right)
- \Tr \left( \raux \paira \pairc \overline{\rho}_{0,\a}  \right) 
- \Tr \left( \raux \overline{\rho}_{0,\a} \paira \pairc \right)
\label{eq:dissipator_trace_pumps}
\end{equation}
We decoupled the trace in Eq.~\eqref{eq:dissipator_trace_pumps}
in both normal and anomalous channels. The three contributions 
become
\begin{equation}
\begin{split}
\Tr &\left( \raux \pairc \overline{\rho}_{0,\a} \paira \right) =  
\Tr \left( \overline{\rho}_{0,\a} \ca_{i \down} \cc_{i \down} \right) 
\Tr \left( \overline{\rho}_{0,\a} \ca_{i \up} \raux \cc_{i \up} \right)+
\Tr \left( \overline{\rho}_{0,\a} \ca_{i \up} \cc_{i \up} \right) 
\Tr \left( \overline{\rho}_{0,\a} \ca_{i \down} \raux \cc_{i \down} \right) \\
&\phantom{=\sum_i}+ 
\Tr \left( \overline{\rho}_{0,\a} \paira \right) 
\Tr \left( \overline{\rho}_{0,\a} \raux \pairc   \right) +
\Tr \left( \overline{\rho}_{0,\a} \pairc \right) 
\Tr \left( \overline{\rho}_{0,\a} \paira \raux  \right) \\
& = 
\left(1 -\frac{n}{2} \right)
\Tr \left(  \raux \cc_{i \up} \overline{\rho}_{0,\a} \ca_{i \up} \right)
+\left(1 -\frac{n}{2} \right)
\Tr \left( \raux \cc_{i \down} \overline{\rho}_{0,\a} \ca_{i \down}  \right)
+\D^* \Tr \left(  \raux \pairc \overline{\rho}_{0,\a}  \right)
+\D  \Tr \left(  \raux \overline{\rho}_{0,\a} \paira \right)
\end{split}
\label{eq:tr_dissipator1}
\end{equation}

\begin{equation}
\begin{split}
\Tr& \left( \raux \paira \pairc \overline{\rho}_{0,\a}  \right) = 
\Tr \left( \overline{\rho}_{0,\a} \ca_{i \down} \cc_{i \down} \right) 
\Tr \left( \overline{\rho}_{0,\a} \raux \ca_{i \up} \cc_{i \up}  \right) +
\Tr \left( \overline{\rho}_{0,\a} \ca_{i \up} \cc_{i \up} \right) 
\Tr \left( \overline{\rho}_{0,\a} \raux \ca_{i \down} \cc_{i \down}  \right)  \\
&\phantom{=\sum_i}+ 
\Tr \left( \overline{\rho}_{0,\a} \paira \right) 
\Tr \left( \overline{\rho}_{0,\a} \raux \pairc \right) +
\Tr \left( \overline{\rho}_{0,\a} \pairc \right) 
\Tr \left( \overline{\rho}_{0,\a} \raux \paira \right)\\
& = 
\left(1 -\frac{n}{2} \right) \Tr \left( \raux \ca_{i \up} \cc_{i \up} \overline{\rho}_{0,\a}   \right)
+
\left(1 -\frac{n}{2} \right) \Tr \left( \raux \ca_{i \down} \cc_{i \down} \overline{\rho}_{0,\a}   \right)
+ \D^* \Tr \left( \raux \pairc \overline{\rho}_{0,\a} \right)
+ \D\Tr \left( \raux \paira \overline{\rho}_{0,\a}  \right)
\end{split}
\label{eq:tr_dissipator2}
\end{equation}

\begin{equation}
\begin{split}
\Tr & \left( \raux \overline{\rho}_{0,\a} \paira \pairc \right) 
= 
\Tr \left(  \overline{\rho}_{0,\a} \ca_{i \down} \cc_{i \down} \right) 
\Tr \left( \overline{\rho}_{0,\a} \ca_{i \up} \cc_{i \up} \raux  \right) 
+
\Tr \left(  \overline{\rho}_{0,\a} \ca_{i \up} \cc_{i \up} \right) 
\Tr \left( \overline{\rho}_{0,\a} \ca_{i \down} \cc_{i \down} \raux  \right)
\\
&\phantom{=\sum_i}+ 
\Tr \left( \overline{\rho}_{0,\a} \paira \right) 
\Tr \left( \overline{\rho}_{0,\a} \pairc \raux  \right) 
+
\Tr \left( \overline{\rho}_{0,\a} \pairc \right) 
\Tr \left( \overline{\rho}_{0,\a} \paira \raux  \right) \\
&=
\left(1-\frac{n}{2} \right) \Tr \left( \raux \overline{\rho}_{0,\a} \cc_{i \up}  \ca_{i \up}  \right)+
\left(1-\frac{n}{2} \right) \Tr \left( \raux \overline{\rho}_{0,\a} \cc_{i \down}  \ca_{i \down}  \right)+
\D^*\Tr \left( \raux \overline{\rho}_{0,\a} \pairc \right)+
\D \Tr \left( \raux \overline{\rho}_{0,\a} \paira  \right)
\end{split}
\label{eq:tr_dissipator3}
\end{equation}
In the above equations $n \equiv \sum_{\s} \Tr\left( \overline{\rho}_{0,\a} \cc_{i \s} \ca_{i \s}  \right) 
$, and we assumed spin-unpolarized density
$\Tr\left( \overline{\rho}_{0,\a} \cc_{i \up} \ca_{i \up}  \right) 
=
\Tr\left( \overline{\rho}_{0,\a} \cc_{i \down} \ca_{i \down}  \right).$
By inserting Eqs.~\eqref{eq:tr_dissipator1}-\eqref{eq:tr_dissipator3} into Eq.~\eqref{eq:dissipator_trace_pumps}, we obtain
\begin{equation}
\begin{split}
\Tr \left( \raux \d {\cal L}_{\a}[\overline{\rho}_{0,\a}] \right) =&  
\left( 1- \frac{n}{2} \right) \Tr \left(\raux \d {\cal L}_{\a}^{\rm 1-p} \right)
+ 2(\a-1) P \sum_i \left[  \D \Tr \left(\raux \overline{\rho}_{0,\a} \paira \right) + 
\D^* \Tr \left(\raux  \pairc \overline{\rho}_{0,\a}\right) \right] \\
&- P \D \Tr \left( \raux \left[ \paira,\overline{\rho}_{0,\a} \right] \right) + 
P \D^* \Tr \left( \raux \left[ \pairc,\overline{\rho}_{0,\a} \right] \right),
\end{split}
\label{eq:trace_disspiator_pump_final}
\end{equation}
where $\d {\cal L}_{\a}^{\rm 1-p}$ represents the hybrid single-particle 
pump dissipator
\begin{equation}
\d{\cal L}_{\a}^{\rm 1-p} = 
2 P \sum_{i\s} \left( \a \cc_{i\s} \overline{\rho}_{0,\a} \ca_{i \s} -\frac{1}{2} 
\left\lbrace \ca_{i \s} \cc_{i \s},\overline{\rho}_{0,\a} \right\rbrace \right).
\end{equation}
Putting all the terms together, the trace in Eq.~\eqref{eq:liouvillian_trace_pumps}
becomes
\begin{equation}
\Tr \left( \raux {\cal L}_{\a}[\overline{\rho}_{0,\a}] \right)
=
\Tr \left( \raux {\cal L}_{BCS,\a}^{\rm pump}[\overline{\rho}_{0,\a}] \right)
= - i \frac{\Tr \left( \raux H_{BCS}(P), \rho_{0,\a} \right)}{\Tr \left( \rho_{0,\a} \right)}
+ 
\frac{\Tr \left( \raux \d {\cal L}_{BCS,\a}^{\rm pump}[\rho_{0,\a}] \right)}{\Tr \left( \rho_{0,\a} \right)}
\label{eq:liouvillian_pumps_trace_aux_normalized}
\end{equation}
with BCS effective unitary Hamiltonian
\begin{equation}
H_{BCS}(P) = H_{BCS,0} -iP \D \sum_i \paira +{\rm h.c.}
\end{equation}
and BCS effective dissipator
\begin{equation}
\d {\cal L}_{BCS,\a}^{\rm pump} = 
\left( 1- \frac{n}{2} \right) \d {\cal L}_{\a}^{\rm 1-p}
+  2(\a-1) P \sum_i \left[  \D {\rho}_{0,\a} \paira  + 
\D^* \Tr \pairc {\rho}_{0,\a} \right].
\end{equation}
Inclusion of two-body loss leads to extra terms that can be 
obtained from Eq.~\eqref{eq:trace_disspiator_pump_final} upon 
particle-hole transformation.
Eventually, variation of with respect to $\raux$ leads to 
the variational dynamics presented in the main text.

\section{Derivation of the equations of motion}
\label{sec:app_eoms}
We detail the calculation of the equations of motions 
for the observables. As before, we show only calculation 
for the pump case and derive the loss contribution from 
particle-hole symmetry.
For a given operator $\hat{O}$, the equations of motions 
for the normalized expectation values 
read
\JoinUp{0.5}{1}{0.5}{1}{XX}
\begin{equation}
\quave{\hat{O}}_{\a}(t) = 
\Tr \left( \overline{\rho}_{0,\a} \hat{O} \right) = 
\frac{\Tr \left( \rho_{0,\a} \hat{O} \right)}{\Tr \left( \rho_{0,\a} \right)}
\qquad \qquad
\partial_t \quave{\hat{O}}_{\a}(t) = 
\Tr \left( \tikzmark{startXX}{{\cal L}_{BCS,\a}^{\rm pump}[\overline{\rho}_{0,\a}]} \tikzmark{endXX}{\hat{O}} \right)
\end{equation}
where, as explained in the Sec.~\ref{sec:eom_obs_normalization}, 
the over the trace indicates that, in order to take into account 
the normalization, the calculation of the trace must include 
only contractions connecting fermionic operators belonging, 
respectively, to the Liouvillian and to the operator $\hat{O}$.
To proceed, we rewrite the Liouvillian  ${\cal L}_{BCS,\a}^{\rm pump}[\overline{\rho}_{0,\a}]$ as
\begin{equation}
{\cal L}_{BCS,\a}^{\rm pump}[\overline{\rho}_{0,\a}] = 
{\cal L}_{BCS,\a=1}^{\rm pump}
[\overline{\rho}_{0,\a}] + 
2 P (\a-1)
\left(1-\frac{n}{2} \right)  \sum_{i \s} \cc_{i \s} \overline{\rho}_{0,\a} \ca_{i \s}
+
2 P (\a-1) \sum_{i} \left[  \D {\rho}_{0,\a} \paira  + \D^* \Tr \pairc {\rho}_{0,\a} \right],
\end{equation}
where ${\cal L}_{BCS,\a=1}^{\rm pump}$ is the BCS dissipator in the full 
Lindblad limit, and split the equations of motions in full 
Lindblad and hybrid, \ie $\a\neq 1$, contributions
\begin{equation}
\partial_t \quave{\hat{O}}_{\a} = \left[\partial_t \quave{\hat{O}}_{\a} \right]_{\rm Lindblad} 
+ (\a-1) \left[\partial_t \quave{\hat{O}}_{\a} \right]_{\rm hybrid}
\end{equation}
In the following, we consider 
$\hat{O}=\cc_{\bk \s} \ca_{\bk \s}$ and $\hat{O} = \cc_{\bk \up} \cc_{-\bk \down}$.
For simplicity, we remove the index $\a$ in the expectation values, 
namely $\D_{\bk} \equiv \quave{\cc_{\bk \up} \cc_{-\bk \down}}_{\a}$ and 
$n_{\bk \s} \equiv \quave{\cc_{\bk \s} \ca_{-\bk \s}}_{\a}$. 
We also use the following symmetries $n_{\bk \s} = n_{-\bk \s}$, 
$n_{\bk \up} = n_{\bk \down}$ and $\D_{\bk}=\D_{-\bk}$.
The Lindblad contributions are obtained by standard commutation relations  
\begin{equation}
\left[\partial_t n_{\bk \s} \right]_{\rm Lindblad}
= - i \Tr \left( \overline{\rho}_{0,\a} \left[ \cc_{\bk \s} \ca_{\bk \s}, H_{BCS}(P) \right] \right)
+ P \left(1-\frac{n}{2} \right) 
\sum_{i\s} \Tr \left( \overline{\rho}_{0,\a} \ca_{i \s} \left[ \cc_{\bk \s} \ca_{\bk \s}, \cc_{i \s} \right]\right) 
+ \Tr \left( \overline{\rho}_{0,\a} \left[ \ca_{i \s} , \cc_{\bk \s} \ca_{\bk \s} \right]  \cc_{i \s} \right)
\end{equation}
and
\begin{equation}
\left[\partial_t \D_{\bk \s} \right]_{\rm Lindblad}
= - i \Tr \left( \overline{\rho}_{0,\a} \left[ \cc_{\bk \up} \cc_{-\bk \down}, H_{BCS}(P) \right] \right)
+ P \left(1-\frac{n}{2} \right) 
\sum_{i\s} \Tr \left( \overline{\rho}_{0,\a} \ca_{i \s} \left[ \cc_{\bk \up} \cc_{-\bk \down}, \cc_{i \s} \right]\right) 
+ \Tr \left( \overline{\rho}_{0,\a} \left[ \ca_{i \s} , \cc_{\bk \up} \cc_{-\bk \down} \right]  \cc_{i \s} \right),
\end{equation}
which, upon defining the complex gap amplitude $\Phi(P) = \left( -|U| + i P \right) \D$, become
\begin{equation}
\left[\partial_t n_{\bk \s} \right]_{\rm Lindblad} = 
- 2 {\rm Im} \left[ \Phi(P) \D_{\bk} \right]
+ 2 P \left( 1- \frac{n}{2} \right)\left( 1 - n_{\bk \s} \right)
\end{equation}
and
\begin{equation}
\left[ \partial_t \D_{\bk} \right]_{\rm Lindblad}
= i 2 \epsilon_{\bk} \D_{\bk} - i 2 \Phi(P) (2 n_{\bk \s} -1) - 2P \left(1 - \frac{n}{2} \right) \D_{\bk}.
\end{equation}
The hybrid contributions read 
%\JoinUp{0.5}{1}{0.5}{1}{a}
\JoinDown{0.5}{-1.0}{1.5}{-1.0}{nkhyb}
\JoinDown{0.5}{-1.0}{1.5}{-1.0}{nkhyb1}
\begin{equation}
\left[ \partial_t n_{\bk \s}\right]_{\rm hybrid} = 
2 P \left(1-\frac{n}{2} \right)  \sum_{i \s'} 
\Tr
\left( 
\tikzmark{startnkhyb}{\underbrace{\cc_{i \s'} \overline{\rho}_{0,\a} \ca_{i \s'}}}
\tikzmark{endnkhyb}{\underbrace{\cc_{\bk \s'} \ca_{\bk \s'}}}
\right)+ 
2 P\sum_{i} \D \Tr \left( 
\tikzmark{startnkhyb1}{\underbrace{\cc_{\bk \s} \ca_{\bk \s}}} 
\tikzmark{endnkhyb1}{\underbrace{ {\rho}_{0,\a} \paira}} \right) 
+ {\rm c.c.}
\end{equation}
\JoinDown{0.5}{-1.0}{1.5}{-1.0}{dkhyb}
\JoinDown{0.5}{-1.0}{1.5}{-1.0}{dkhyb1}
\begin{equation}
\begin{split}
\left[ \partial_t \D_{\bk}\right]_{\rm hybrid} = 
2 P \left(1-\frac{n}{2} \right)  \sum_{i \s} 
\Tr
\left( 
\tikzmark{startdkhyb}{\underbrace{\cc_{i \s} \overline{\rho}_{0,\a} \ca_{i \s}}}
\tikzmark{enddkhyb}{\underbrace{\cc_{\bk \up} \cc_{-\bk \down}}}
\right)+ 
2 P \sum_{i} & \D \Tr \left(
\tikzmark{startdkhyb1}{\underbrace{\cc_{\bk \up} \cc_{-\bk \down} }}
\tikzmark{enddkhyb1}{\underbrace{{\rho}_{0,\a} \paira}} \right) + \\
& \D^* \Tr 
\left( 
\tikzmark{startdkhyb2}{\underbrace{\cc_{\bk \up} \cc_{-\bk \down} }}
\tikzmark{enddkhyb2}{\underbrace{\pairc {\rho}_{0,\a} }}
\right)
\end{split}
\end{equation}
where the lines indicate contractions connecting fermionic operators on the two 
groups of operators.
This can easily be computed with the help of translation and spin symmetry, $\quave{\cc_{\bk \s} \ca_{\bk' \s'}} \propto \d_{\bk \bk'} \d_{\s \s'}$ and $\quave{\cc_{\bk \s} \cc_{\bk' \s'}} \propto \d_{\bk -\bk'} \d_{\s,-\s'}$,
leading to 
\begin{equation}
\left[ \partial_t n_{\bk \s}\right]_{\rm hybrid} = 
2 P \left[  \left( 1 -\frac{n}{2}\right) \left[ (1-n_{\bk \s})^2 -|\D_{\bk}|^2 \right] 
+ 2 {\rm Re}[\D \D_{\bk}^*] (1-n_{\bk \s}) \right]
\end{equation}
and
\begin{equation}
\left[ \partial_t \D_{\bk}\right]_{\rm hybrid} =
2 P \left[  \left( 2-n \right) \D_{\bk} (n_{\bk \s} - 1) + \D (1- n_{\bk \s})^2 - \D^* \D_{\bk}^2 \right]
\end{equation}
The equations of motions for the losses are obtained by applying particle-hole 
transformations $\D_{\bk} \to -\D_{\bk}^*$, $n_{\bk \s} \to (1- n_{\bk \s})$
and leads to the full equations of motion presented in the main text.
 
%\section{Non-hermitian equations of motion for the order }

\end{document}